\documentclass[USenglish,oneside,twocolumn]{article}
\usepackage[utf8]{inputenc}

\usepackage[big]{dgruyter_NEW}
\usepackage[dvipsnames]{xcolor}
\usepackage{graphicx}
\usepackage{adjustbox}
\usepackage{graphics}
\usepackage{comment}
\usepackage{placeins}
\usepackage{float}
\usepackage{amsfonts}
\usepackage{amsthm}
\usepackage[ruled,vlined,linesnumbered]{algorithm2e} 
\usepackage{caption}
\usepackage{subcaption}
\usepackage[shortlabels]{enumitem}
\usepackage{xspace}
\usepackage{rotating}
\usepackage{pdflscape}
\usepackage{float}
\theoremstyle{definition}
\newtheorem{definition}{Definition}[section]

\newtheorem{theorem}{Theorem}[section]

\newcommand{\our}{DP-Loc\xspace}

\usepackage[utf8]{inputenc}

\DOI{foobar}

\cclogo{\includegraphics{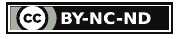}}
  
\newcommand{\red}[1]{{\color{red}{#1}}}  
\newcommand{\rev}[1]{{\color{black}{#1}}}  
\newcommand{\revs}[1]{{\color{black}{#1}}}
\newcommand{\revss}[1]{{\color{black}{#1}}}
  
\begin{document}

  \author*[1]{Szilvia Lestyán}

  \author[2]{Gergely Ács}

  \author[3]{Gergely Biczók}

  \affil[1]{Department of Networked Systems and Services,CrySyS Lab, Budapest University of Technology and Economics, E-mail: lestyan@crysys.hu}

  \affil[2]{E-mail: acs@crysys.hu}

  \affil[3]{E-mail: biczok@crysys.hu}

  \title{\huge In Search of Lost Utility: Private Location Data}

  \runningtitle{In Search of Lost Utility: Private Location Data}


  \begin{abstract}
  {
	The unavailability of training data is a permanent source of much frustration in research, especially when it is due to privacy concerns. This is particularly true for location data since previous techniques all suffer from the inherent sparseness and high dimensionality of location trajectories which render most techniques impractical, resulting in unrealistic traces and unscalable methods. Moreover, time information of location visits is usually dropped, or its resolution is drastically reduced.
	In this paper we present a novel technique for privately releasing a composite generative model and whole high-dimensional location datasets with detailed time information. To generate high-fidelity synthetic data, we leverage several peculiarities of vehicular  mobility such as its language-like characteristics (``you should know a \textit{location} by the company it keeps'') 
	or how humans plan their trips from one point to the other. 
	We model the generator distribution of the dataset by first constructing a variational autoencoder to generate the source and destination locations, and the corresponding timing of trajectories. Next, we compute transition probabilities between locations with a feed forward network, and build a transition graph from the output of this model, which approximates the distribution of all paths between the source and destination (at a given time). Finally, a path is sampled from this distribution with a Markov Chain Monte Carlo method.
	The generated synthetic dataset is highly realistic, scalable, provides good utility and, nonetheless, provably private. We evaluate our model against two state-of-the-art methods and \rev{three} real-life datasets demonstrating the benefits of our approach \footnote{\revss{The research was supported by the Ministry of Innovation and Technology NRDI Office within the framework of the Artificial Intelligence National Laboratory Program.}}.
	}
    \end{abstract}


  \journalname{Proceedings on Privacy Enhancing Technologies}
\DOI{Editor to enter DOI}
  \startpage{1}
  \received{..}
  \revised{..}
  \accepted{..}

  \journalyear{..}
  \journalvolume{..}
  \journalissue{..}

\maketitle
	
	\section{Introduction}\label{intro}

	Analyzing human mobility patterns has been in the focus of both researchers and practitioners in the last decades \cite{gonzalez2008understanding} \cite{biczok2014navigating} \cite{intelligent_trans}.
	\rev{Having created an already \$12 billion market, the location data industry is steadily expanding with companies that harvest, sell, or trade in location data. Many of these firms claim that privacy has utmost importance in their businesses and that they never sell personal data.}\footnote{\url{https://themarkup.org/privacy/2021/09/30/theres-a-multibillion-dollar-market-for-your-phones-location-data}}
	
	However, collecting and mining location data come inherently with their own strong privacy and other ethical concerns~\cite{beresford2003location}.
	Several studies have shown that pseudonymization and quasi-standard de-identification are not sufficient to prevent users from being re-identified in location datasets \cite{de2013unique4points} \cite{zang2011anonymization}. This significantly hinders location data sharing and use by researchers, developers and humanitarian workers alike\footnote{\url{https://www.economist.com/leaders/2014/10/23/call-for-help}}.
	
	Although a plethora of different anonymization techniques have been proposed for location data (for a very thorough survey see \cite{fiore2019trajsurvey}), they all suffer from either weak utility, or weak privacy guarantees, or they are not scalable to large datasets.
	Indeed, location data are inherently high-dimensional and often sparse, which makes an individual's location trajectory unique even in very large populations\footnote{Four data points—approximate places and times where an individual was present—have been demonstrated to be enough to uniquely re-identify 95\% of the users in a dataset of 1.5 million users \cite{de2013unique4points}}. This has a detrimental effect on privacy, and also on utility due to the curse of dimensionality. Note that  aggregation per se does not necessarily prevent such re-identification in practice \cite{XuTLZFJ17, PyrgelisTC18}. 
	The brittleness of privacy guarantees ignited research towards location anonymization with provable privacy guarantees. So far, only a handful of prior works \cite{chen2012diffgrams,he2015dpt,gursoy2018utility, SGLT} have addressed the off-line anonymization of complete location trajectories with formal privacy guarantees. Most of these approaches use some form of Differential Privacy \cite{dwork2006calibrating}, which has become the \textit{de facto} privacy model recent years \cite{ErlingssonPK14, Abowd18,appleDP}. 
	
	Unfortunately, off-line location anonymization with Differential Privacy often implies serious accuracy degradation which can make anonymized data useless in practice. In particular, most schemes follow the common three-steps approach to generate privacy preserving synthetic location traces: (1) finding a faithful generative model of the underlying data generating distribution, (2) adding noise to the training of this model in order to provide Differential Privacy, (3) generating synthetic location trajectories from the noisy generative model. Generally, inaccuracy stems from either using a sub-optimal generative model, or a model which is not sufficiently robust against the additional noise needed for Differential Privacy. Although a more complex model is capable of faithfully capturing the peculiarities of location data, it also requires larger perturbation for privacy owing to the increased number of its parameters. Moreover, complex models very often are  not scalable to larger datasets with several hundreds of thousands of location trajectories \cite{SGLT,he2015dpt}. Finally, to the best of our knowledge, none of these differentially private anonymization approaches release the valuable fine-grained temporal characteristics of location visits.   
	
	In this paper, we propose a novel off-line anonymization scheme called \our for location data with strong Differential Privacy guarantees. Unlike prior works, we tackle high dimensional data modeling with generative neural networks (GNN) which have shown great promise recently. GNNs have the potential to automatically learn the general features of a location dataset including complex regularities such as the subtle and valuable correlation among different location visits as a function of time.
	The key of our approach is a novel decomposition of the  generator model into a sequence of smaller models and a post-processing step, which are robust against perturbation needed for Differential Privacy, and therefore can be used to generate high-fidelity synthetic location trajectories. In particular, we first project all trajectories to a smaller set of frequent locations thereby reducing the dimensionality of the data. Then, a synthetic trace is created by first generating its source and destination along with time information, and then finding a path on a transition graph between these endpoints. Both the endpoint and graph generation are modelled by distinct neural networks scalable for being trained with differentially private gradient descent \cite{abadi2016deep}.
	
	\our has several advantages. First, projecting traces to a smaller set of locations helps generalization and also increases robustness against the added noise. Second, separating endpoint from path generation preserves the length of trajectories more accurately compared to related work. Third, recently proposed advanced composition theorems of Differential Privacy \cite{abadi2016deep} enable us to use differentially private neural networks with better utility instead of simple Markovian models \cite{gursoy2018utility,chen2012diffgrams}, and quantify the privacy guarantee of \our more accurately.
	Finally, neural networks are sufficiently flexible to model transition probabilities depending on the destination and time, which facilitates the generation of more realistic location traces. Figure \ref{fig:heatmaps} illustrates the power of \our, where the density of complete synthetic location trajectories produced by \our are compared to the original data and a state-of-the-art solution from \cite{gursoy2018utility}.

	\noindent Our specific contributions are as follows:
	\begin{enumerate}[1.,topsep=0pt]
		\item We propose a novel generative model, \our to generate realistic location trajectories with time information. A Variational Auto-Encoder (VAE) is used to generate the source and destination of a trace along with a single timestamp of the trace. Then, a feed-forward neural network is built to compute the transition probabilities between any two locations depending on the destination and time, which implicitly defines the distribution of all paths between the source and the destination at a given time. Finally, a realistic path is generated by sampling from this distribution with a Markov Chain Monte Carlo (MCMC) method. 
		\item  Our composite generative model can be trained with differentially private gradient descent \cite{abadi2016deep} on sensitive location data, and, therefore, can be used to synthesize the training data with formally proven privacy guarantees. Such synthetic training data can be shared for any purposes without violating individuals' privacy.
		\item We evaluate our model on \rev{three} real-life public mobility datasets, and demonstrate that the generated private synthetic data has higher utility compared to previous works.
		
	\end{enumerate}

	\section{Related work}
	\label{sec:related}
	A prominent line of anonymization research uses the notion of Differential Privacy \cite{dwork2008differential} which gives a privacy guarantee based on rigorous mathematical proofs. Some proposals based on synthetic data generation via machine learning, i.e., modeling the dataset through the underlying distributions of generating variables, do apply Differential Privacy specifically for location data~\cite{chen2012diffgrams,he2015dpt}, but with significant shortcomings. He et al.\cite{he2015dpt} (DPT) discretize raw GPS trajectories using hierarchical reference systems to capture individual movements at different speeds. They then propose an adaptive mechanism to select a small set of reference systems to construct prefix tree counts. Lastly, a direction-weighted sampling is applied to improve utility.

	 Chen et al. designed Ngram \cite{chen2012diffgrams}, a variable-length n-gram model that makes use of an exploration tree structure and Markovian assumptions, with Differential Privacy.
	 Gursoy et al. \cite{gursoy2018utility} designed AdaTrace, a generative model with a four-phase synthesis process consisting of feature extraction, synopsis learning, differentially private noise injection and synthetic trace generation. Additionally, they provided defense against three ad-hoc privacy attacks.
	 Bindschaedler and Shokri \cite{SGLT} (SGLT) enforce plausible deniability to generate privacy-preserving synthetic traces. SGLT first recommends trace similarity and intersection functions that map a synthetic trace to an original one under similarity and intersection constraints. Then, it generates one synthetic trace using one real trace as its seed. If the synthetic trace satisfies plausible deniability, i.e.,  there exist $k$ other real traces that can be mapped to the synthetic trace, then it preserves the privacy of the seed trace.
     DPT and Ngram generate synthetic traces as random walks that do not incorporate destinations. However, a realistic trajectory heavily depends on the destination, and humans rarely visit a location following a \textit{``let's go to a place where people usually go from here''} policy. In addition, a random walk can possibly generate synthetic traces with unrealistic length (see details in Section \ref{sec:experimental}).  Moreover, time-of-day is also left out from the models (Ngram, DPT, AdaTrace); clearly, this reduces the descriptive power of the generative model as human mobility does show strong time-of-day patterns~\cite{gonzalez2008understanding}. In fact, time-of-day even influences trip destinations. (Just consider how your destination varies from 8am to 8pm.) Simply dropping the timestamps averages out the visit frequencies between night and day, morning and afternoon etc. SGLT applies large bins in order to incorporate time (morning, afternoon, evening and night). However, SGLT is mainly designed for CDR (Call Detail Record) datasets, and it utilizes the peculiarities of the implied mobility patterns, such as semantic similarity (SGLT) between locations. This approach can hardly generate valuable synthetic data when it comes to short term, dense movements, such as GPS trajectories of vehicles between start and end locations. Borrowing the example from \cite{SGLT}, consider Alice and Bob spending all day at their respective work locations $w_A$ and $w_B$, and all night at their respective home locations $h_A$ and $h_B$. Obviously, their mobility models are semantically very similar, although it might be the case that $h_A\neq h_B$ and $w_A\neq w_B$. In this example, the best semantic mapping between locations will be $w_A \leftrightarrow w_B$ and $h_A \leftrightarrow h_B$. This example clearly introduces the type of mobility traces generated by this model. In our experiments, we applied datasets of vehicle trajectories where the above illustrated semantic similarity is not applicable at all times.
     \rev{Furthermore, in CDR-like data, we can mostly observe the locations where people reside for longer periods (such as work and home), while \our models individuals' movements from one of these locations to another, where the turns and stops depend on the hour of the day (e.g., owing to traffic jams). For example, oftentimes we take a different route from home to work depending on the traffic conditions. More formally, in the case of CDR (or similar) datasets, the transition probabilities between locations are closer to uniform; however, when we increase the sampling rate of locations along a route, this does not hold anymore (particularly when we also condition on time and destination), thus allowing us to build better models.}
     \revs{Our model is best suited for trips that have a start and destination that can be fit in the same time-slot. For different datasets different time granularity can be used based on the underlying application. However, for CDR-like data, we suggest two possible alterations: (1) setting one time-slot to one day, or (2) remove all time information. 
     The evaluation of these scenarios go beyond the scope of this paper, hence we leave them for future research.}

     Ngram, DPT and AdaTrace apply the Laplace mechanism to preserve Differential Privacy, where the noise is added to the Markovian probabilities. In contrast to these, we apply the Moments Accountant \cite{abadi2016deep} method with the Gaussian mechanism that utilizes an advanced composition theorem by taking into account the exact noise distribution.
     Finally, SGLT has been shown \cite{gursoy2018utility} to be very slow, while DPT requires large computational power  (256GB RAM and 48 cores for $50,000$ traces, whereas we generate $400,000$ traces), thus these models are not scalable for large datasets.
     
    Most privacy-preserving training algorithms for neural networks are based on modifying the gradient information generated during backpropagation. The modification involves clipping the gradients (to bound the influence of any single record on the released model parameters) and adding calibrated random noise \cite{abadi2016deep, chen2020gswgan}. Some works propose to use generative adversarial networks (GANs) \cite{park2018data} or mixture models \cite{acs2018diffmixture} to directly generate privacy-preserving synthetic data. 
    \rev{Differentially Private GAN \cite{XieGANDP, ZhangGANDP, DPCGAN} has been applied to generate image and \revs{Electronic Health Record} data.
    The approach in \cite{FrigerioOGD19} aims to generate time series with \revs{LSTM (Long-Short Term Memory Networks) and GAN (Generative Adversarial Networks)} with DP guarantees, as well as multi-variate tabular data.
    GS-WGAN \cite{ChenOF20} uses gradient-sanitized Wasserstein GAN to generate synthetic data and has also been demonstrated on image data.
    Another DP-GAN architecture was proposed in \cite{Beaulieu-Jones159756} to release patient-level clinical trial data with Differential Privacy.
    All these techniques use some variant of DP-SGD \revs{(Differentially Private Stochastic Gradient Descent)} \cite{abadi2016deep} for training the discriminator of GAN. By contrast, PATE-GAN \cite{JordonYS19a} was proposed to generate synthetic multi-variate tabular data using PATE \cite{PapernotAEGT17}.}
    Nonetheless, none of these generative models are specific to location data generation.
	\vspace{-.3cm}
	\section{Preliminaries}
	\label{sec:pre}
	
	\subsection{Location Data} \label{locdata}
	In general, location data is geographical information about a specific object’s whereabouts associated to a time identifier. 
	Formally, let $\mathbf{L} = \{L_1 , L_2 ,\ldots , L_{|\mathbf{L}|} \}$ be the universe of locations, where $|\mathbf{L}|$ is the size of the universe. We assume that the whole universe is represented as a grid, and each location corresponds to a cell in the grid.
	Each record in a location database is a sequence of timestamped location visits drawn from the universe. Specifically, a sequence $S$ of length $|S|$ is an ordered list of items $S = (L_{\ell_1},t_1) \to (L_{\ell_2}, t_2) \to \ldots \to (L_{\ell_{|S|}},t_{|S|})$, where $\forall 1 \leq i \leq |S|,\: L_{\ell_i} \in
	\mathbf{L}$. A location may occur multiple times in $S$. A location database $D$ is composed of a multiset of sequences $D = \{S_1 , S_2 ,\ldots , S_{|D|} \}$, where $|D|=N$ denotes the number of traces in $D$. \\
	
	\vspace{-.4cm}	
	\subsection{Differential Privacy}\label{dp}
	
	Differential Privacy \cite{dwork2006calibrating} (DP) ensures that the outcome of any computation on a database  is insensitive to the change of a single record. It follows that any information that can be learned from the database with a record can also be learned from the one without that particular record. In our case,  DP guarantees that our generative model is not affected by any single original trajectory beyond the privacy budget measured by $\varepsilon$ and $\delta$, which can be computed as follows.
		
		\rev{\begin{definition} [\textbf{$(\epsilon, \delta)$-Differential Privacy} \cite{dwork2006eddiff}]
		\label{def:DP} A privacy mechanism $\mathcal{M}$ gives $(\epsilon, \delta)$-Differential Privacy if for any database $D_1$ and $D_2$ differing on at most one record and for any subset of outputs $S \subseteq \mathit{Range}(\mathcal{M})$:
	   $$\Pr[\mathcal{M}(D_1) \in S] \leq e^{\varepsilon} \Pr[\mathcal{M}(D_2) \in S] + \delta$$
	   
\end{definition}

Intuitively, a privacy mechanism $\mathcal{M}$ satisfying Definition~\ref{def:DP} does not release any information that is specific to any single record in dataset $D$ up to $\varepsilon$ and $\delta$, where $\delta$ is preferably smaller than $1/|D|$ \cite{dwork2006eddiff}.}

	
	A fundamental concept for achieving Differential Privacy is the global sensitivity of a function \cite{dwork2006calibrating}:
	
	\begin{definition}[\textbf{Global $L_p$-sensitivity} \cite{dwork2014algorithmic}]
		For any function $f : \mathcal{D}\to R^d$ , the $L_p$-sensitivity of $f$ is
			$\Delta f = \max_{\mathcal{D}_1, \mathcal{D}_2} || f(D_1) - f(D_2)||_p$
		for all $D_1, D_2$ differing in at most one record, where $||\cdot||_p$ denotes the $L_p$-norm.
	\end{definition}
	
	Differential Privacy maintains composition; the privacy guarantee of the  $k$-fold adaptive composition  of any mechanism can be computed using the moments accountant method \cite{abadi2016deep}. 
    The moments accountant  generalizes the regular approach of
	keeping track of $(\epsilon, \delta)$ using an advanced composition theorem by 
	taking into account the exact noise distribution. 

	\begin{definition}[\textbf{Privacy loss} \cite{abadi2016deep}]
		Let $\mathcal{M}$ be a privacy mechanism which assigns a value $O \in Range(\mathcal{M})$ to a dataset $D$. The privacy loss of $\mathcal{M}$ with datasets $D_1$ and $D_2$ and auxiliary input $aux$ at output $O$ is a random variable:
		$$\mathcal{L}(O;D_1, D_2, \mathcal{M}, aux)=\log \frac{P[\mathcal{M}(D_1, aux)=O]}{P[\mathcal{M}(D_2, aux)=O]}$$
	\end{definition}
	
	\begin{definition}(\textbf{Log of the Moment Generating Function})
	\label{def:moments} For a given mechanism $\mathcal{M}$, the log of the moment generating function evaluated at $\lambda$ is:
		$\alpha_{\mathcal{M}}(\lambda; aux, D_1, D_2) = \log\: \mathbb{E}_{O\sim \mathcal{M}(D_1, aux)} [\exp(\lambda\mathcal{L}(O; D_1, D_2, \mathcal{M}, aux))].$
	\end{definition}
	
	\begin{theorem} \label{Moments} 
	(\textbf{Moments Accountant} \cite{abadi2016deep})
		Let \rev{$\alpha_\mathcal{M}(\lambda) = \max_{aux, D_1, D_2} \alpha_{\mathcal{M}}(\lambda; aux, D_1, D_2)$} be defined as above.
		Let $\mathcal{M}_{1:k}$ be the $k$-fold adaptive composition of $\mathcal{M}_1,\ldots , \mathcal{M}_k$. Then:
		\begin{enumerate}
			\item \textit{Composability:} $\alpha_{\mathcal{M}_{1:k}}(\rev{\lambda}) \leq \sum_{i=1}^k \alpha_{\mathcal{M}_i}(\lambda)$
			\item \textit{Tail bound:} For any $\varepsilon > 0$, the mechanism $\mathcal{M}_{1:k}$ is $(\epsilon, \delta)$-differentially private for $\delta=\min_{\lambda}\: \exp(\alpha_{\mathcal{M}_{1:k}}(\lambda)- \lambda\epsilon)$.
		\end{enumerate}
	\end{theorem}
	
     The \textit{Gaussian Mechanism} \cite{dwork2014algorithmic} consists of adding Gaussian noise to the true output of a function. In particular, for any function $f:D\to R^n$, the Gaussian mechanism is defined as adding i.i.d. Gaussian noise with variance $(\Delta_2f \cdot \sigma)$ and zero mean to each coordinate value of $f(D)$.
	In fact, the Gaussian mechanism draws vector values from a multivariate isotropic Gaussian distribution
	which is described by random variable $\mathcal{G}(f(D),\Delta_2 f\cdot\sigma \mathbf{I}_n)$.

	\section{Model}
	\label{sec:model}

	\subsection{Overview}
	Our goal is to generate private synthetic location traces. 
	In particular, having a location dataset $D$ with a multiset of trajectories, our goal is to build a generative model which approximates the true generator distribution of $D$, where every trajectory in $D$ is a sample from this distribution. The model is built using the privacy-sensitive data $D$, and, hence, the training process of this model must guarantee Differential Privacy for any user/trajectory in $D$.
	Due to the large complexity of this model, we decompose it into four main parts as follows: 
	\begin{enumerate}[-,topsep=0pt]
	    \item \textbf{Dimensionality reduction:}	We identify the $K$ most frequently visited cells on the map and work only with these locations afterwards by projecting all trajectories to these cells.
		\item \textbf{Trajectory Initialization:} A generative model, called \emph{Trajectory Initializer (TI)}, learns the underlying joint distribution of the starting and ending locations and time variable of all trajectories, i.e., their very first (source) and very last (destination) location visits along with the single timestamp of the whole trace. 
		\item \textbf{Transition Probability Generation:} A classification model, called \emph{Transition Probability Generator (TPG)}, learns the transition probability distribution between any two consecutive locations, i.e., it outputs the probability distribution for the next hop in a trace, conditioned on the current location, the destination and time. 
		Both of these models are trained with Differential Privacy guarantees on a potentially sensitive training dataset. 
		\item \textbf{Trace Generation:} Sampling a source $L_{src}$ and destination $L_{dst}$ along with the time $t$ from the output distribution of TI, and using the transition probabilities between any locations generated by TPG, the trace generator (TG) non-deterministically reconstructs a trajectory between source $L_{src}$ and destination $L_{dst}$ at time $t$ combining Dijkstra's shortest path and the Metropolis-Hastings algorithm.
		As this process only uses the output of TI and TPG, and some public information about locations, the whole generation process becomes differentially private as the first two are already differentially private.
\end{enumerate}
	\smallskip

\noindent \textbf{Assumptions:} Time is divided into equally sized slots which are sufficiently large to include whole trajectories. Each trajectory is assigned to a single time slot $t$, and all location visits of a trajectory take place within this slot $t$. 
		
	\begin{algorithm}[t]
		\small
		\caption{Differentially Private Synthetic Trace Generator (\our)\label{alg:ours}}
		\DontPrintSemicolon
		\KwIn{Private Dataset $D$}
		{\bf Dimensionality reduction:}\;
		Choose $K$ most frequent locations TOP-$K$: $L_{\ell_1},...L_{\ell_K}$, $\forall L_{\ell_i}\in \mathbf{L}$ with Gaussian Mechanism\;
		Project each trace $p \in D$ to TOP-$K$\;
		{\bf Model construction:}\;
		Train the Trace Initialization Model $TI_{\theta_1}$ on $D$ with DP-SGD \cite{abadi2016deep}\; 
		Train the Transition Probability Generation Model $TPG_{\theta_2}$ on $D$ with DP-SGD \cite{abadi2016deep}\; 
		\textbf{Trajectory Reconstruction:}\; 
		\For {$i \in [1, \ldots, |D|]$}
		{
			Sample $(L_{src}, L_{dst}, t) \sim TI_{\theta_1}$\;
			Build a routing graph $G(V,E)$, where $V= \mathbf{L}$ and $weight((L_x, L_y)) = -\log TPG_{\theta_2}[L_y | t, L_x, L_{dst}]$, where $(L_x, L_y) \in E$\;
			Find the path $p$ between $L_{src}$ and $L_{dst}$ with the minimal total weight in $G$\;
			\For {$\rev{m  \in [1, \ldots, 10]}$}
			{
			\rev{Run Metropolis-Hastings on $p$: $\revs{p} = MH(p)$} \;
			}
			Generate repetitions in the sampled path $s =(L_{\ell_1}, \ldots, L_{\ell_n})$:\;
			\For{$L_{\ell_j} \in s$}
			{
				$\eta \sim  Geom(1-TPG_{\theta_2}[L_{\ell_j}| t, \rev{L_{\ell_{j-1}}}, \rev{L_{\ell_n}}])$\;
				$s' = (L_{\ell_1}, \dots, L_{\ell_{j-1}}, \underbrace{L_{\ell_j},\dots, L_{\ell_j}}_{\text{$\eta$ times}}, L_{\ell_{j+1}}, \dots, L_{\ell_n}) $
			}
		
			$D' = D' \cup \{(s', t)\}$\;
		}
		
		\KwOut{Synthetic dataset $D'$} 

	\end{algorithm}

	\subsection{Model Description}\label{modeldesc}
	In the remainder of this section, we describe our approach called \our (Differentially Private Synthetic Trace Generator) in more details that is also summarized in  Alg.~\ref{alg:ours}.
	
	\subsubsection{Dimensionality reduction}
	\label{sec:dimred}
    We project all locations of every trajectory to the $K$ most frequently visited locations (cells) on the map called as TOP-$K$ locations. In particular, each location is mapped to the closest TOP-$K$ location if there is any within a distance of 1000 meters. Otherwise, the whole trace is dropped. The value $K$ \rev{refers to the number of most frequently visited cells where 95\% of all visits occur; thus, it differs among datasets, and must be chosen in a differentially private fashion (see Section \ref{results} and Section \ref{DP_params}). (We made an exception and lowered the threshold percentage in case of GeoLife-250 dataset to 80\% because of the low amount of datapoints in many cells. The same preprocessing was done for Ngram and AdaTrace as well.)}
    The purpose of this dimensionality reduction is to increase model accuracy and also the speed of training. Indeed, if there are many cells on the map, the number of visits per cell typically has a power-law distribution: most cells are never or rarely visited. 
    Using all grid cells 
    would largely increase model complexity and hence training time. Moreover, it also degrades model quality due to the larger perturbation needed by DP (see Section \ref{DP_params}). \revss{On the other hand, dimensionality reduction helps the model preserve large-scale mobility patterns more accurately (with high support) at the expense of losing fine-grained patterns (with low support).} This trade-off can be dynamically chosen per application; we believe that retaining 95\% of all visits provides reasonably high fidelity.
    
	
	\vspace{-.2cm}
	\subsubsection{Trajectory Initialization (TI)}\label{VAE}
	In order to sample a starting location $L_{src} \in \mathbf{L}$, a destination $L_{dst} \in \mathbf{L}$ and time  $t \in \mathbf{T}$ for a synthetic trajectory, we build a differentially private Variational Autoencoder (VAE) (see Figure \ref{fig:vae} in Appendix \ref{app:complexity} for illustration) that is capable of approximating the joint probability distribution $Pr(L_{src}, L_{dst}, t)$.
	The model parameters $\theta_1$ are learnt from a sensitive location dataset $D$, and hence training is performed with DP-SGD \cite{abadi2016deep} (see Section \ref{DP_params} for details).
	
	The output of $TI_{\theta_1}$ is a 3-dimensional vector $[L_{src}, L_{dst}, t]$ (recall that each trace has a single timestamp in our model).
	Notice that learning this distribution privately is challenging due to its high dimensionality; the domain of the joint probability distribution is $|\mathbf{L}| \times |\mathbf{L}| \times |\mathbf{T}|$, where $|\mathbf{T}|$ is the number of all possible time slots. 
	
	We one-hot encode the input, thus it has a dimension of $2\times|\mathbf{L}|+24$, where the size of $|\mathbf{L}|$ depends on the coarseness of the grid, and $24$ is the number of hours in a day. 
	Our encoder has two hidden \textit{dense} layers $(100,100)$ with ReLU and linear activation functions, respectively. The encoder outputs the parameters of the learned normal distribution $\mathcal{N}(\mu, \sigma)$; values drawn from this distribution by the decoder comprise the latent vectors of size $50$.
	The decoder has to transform this latent variable to an actual sample. The decoder has only $1$ hidden layer with size $100$ and with ReLU activation. Finally, there are three parallel output layers with softmax activation, corresponding to a single output variable (location, destination, time). VAEs have their own specific loss functions; we applied the original one from \cite{doersch2016tutorialVAE}. 

	
	\vspace{-.2cm}
	\subsubsection{Transition Probability Generation (TPG)}
	\label{FFN}
	Our classifier $TPG_{\theta_2}$ is a feed forward network (FFN) endowed with word embedding. It is illustrated in Figure \ref{fig:tpg} in the Appendix. It approximates the true transition distribution $Pr[L_{x} \rightarrow L_{y} | t, L_{dst}]$ for any frequent (see details below) location $L_{x}$ and $L_{y}$ for every possible time slot $t \in \mathbf{T}$ and destination $L_{dst} \in \mathbf{L}$. That is, the probability that an individual at location $L_{x}$ moves to location $L_{y}$ towards destination $L_{dst}$ at time $t$.
	
	The input is $(L_{c},L_{dst},t)$ (current location, destination, time), and the output is the probability distribution on the next hop. The two location coordinates of the input vector are fed into an embedding layer where they are embedded separately into the same $50$-dimensional vector space\footnote{The embedding layer is part of the network, thus trained together with the rest of the layers.}. Next, we concatenate these vectors with the time coordinate, resulting in a $101$-dimensional vector. The next dense layer has a size of $200$ with ReLU activation, and the output layer has softmax. We trained the network with \textit{sparse categorical cross-entropy} and the \textit{SGD} optimizer. As only the K most frequently visited cells are considered,  the number of output classes is also $K$.
	We use DP-SGD \cite{abadi2016deep} to train TPG and therefore the released model parameters $\theta_2$ are differentially private (see Section \ref{DP_params} for details).
	Besides the current time and destination, the prediction of the next location depends only on the current location and not on the earlier location visits. That is, when the next location is predicted,  we do not take into account how the current location is reached. This is not a far-fetched simplification; several studies have shown that 1 or at most 2-order Markov chains provide a sufficiently accurate estimation of the next location visit \cite{gambs2012nextmarkov}. 

	\subsubsection{Remarks}
	\rev{The choice of an embedding layer, and an FFN instead of a recurrent neural network (such as LSTM) deserves more explanation.}
	As locations exhibit similar characteristics to words, we can rely on the distributional hypothesis\footnote{\rev{Words that co-occur in the same contexts tend to have similar meanings}}. \rev{In case of locations, this means that if they follow each other in a given trajectory, then they are also close in the geographical space, and, therefore, will have similar representations in the embedded space. This implies that the model can automatically learn which locations are geographically close to each other.}
	
	\rev{Although LSTM has an implicit capability to handle temporal and sequential data}, its differentially private training with SGD takes approximately 6-8 times longer than  without Differential Privacy. However, training simple feed forward networks is considerably faster, almost as fast as the non-private model. Furthermore, our FFN has less parameters than the simplest but still well-performing LSTM, and having less parameters results in lower noise injection, and thus, higher utility. Our solution has an accuracy only 1-2$\%$ less than the LSTM layer on the considered datasets.

    \vspace{-.2cm}
	\subsubsection{Trace Generation (TG)}\label{TG}
	When a trajectory is generated, we first sample a pair of source $L_{src}$ and destination $L_{dst}$ locations along with the time slot $t$ from the output distribution of $TI_{\theta_1}$. Then, a weighted directed routing graph $G(V, E)$ is built, where the edge weights are the transition probabilities between any two location points generated by $TPG_{\theta_2}$ (i.e., $V$ is composed of locations in TOP-K, and $weight((L_{x}, L_{y})) = -\log Pr[L_{x} \rightarrow L_{y} | t, L_{dst}]$ for any $(L_x, L_y) \in E$, that is the negative logarithm of the transition probability from $L_x$ to $L_y$ conditioned on destination $D$ and time $t$). Note that $G$ is complete and specific to a given destination $L_{dst}$ and time slot $t$, hence different graphs are constructed for trajectories differing in their destination or time. The routing graph defines a distribution of paths between any location and destination $L_{dst}$ at time $t$, and our task is to draw a path from this distribution in order to generate a trajectory.
	
	To do so, the most probable trajectory is first constructed from graph $G$ by applying Dijkstra's shortest path algorithm, and then the Metropolis-Hastings MCMC algorithm to the resulting shortest path, thus we generate one of the most probable paths between $L_{src}$ and $L_{dst}$.
	As $-\log TPG_{\theta_2}[L_y | t, L_x, L_{dst}]$ is always non-negative, Dijsktra's shortest path algorithm finds the path with the minimum total weight between two vertices, which is equivalent to the most probable path between $L_{src}$ and $L_{dst}$ at time $t$ in our case. Indeed, let $P$ denote the set of all paths between $L_{src}$ and $L_{dst}$. Then, the most probable path between $L_{src}$ and $L_{dst}$ is
	{\small
	\begin{align*}
	&\min_{p \in P} \sum_{(L_x \to L_y) \in p} - \log(TPG_{\theta_2}[L_{y} | t, L_{x}, L_{dst}])\\
	&=\min_{p\in P} - \log(\prod_{(L_x \to L_y) \in p} TPG_{\theta_2}[L_y | t, L_x, L_{dst}]) \\
	&\approx\min_{p\in P} - \log(\prod_{(L_x \to L_y) \in p} Pr[L_x \to L_y | t, L_{dst}]) \\
	&=\max_{p \in P} \prod_{(L_x \to L_y) \in p} Pr[L_x \to L_y | t, L_{dst}] 
	\end{align*}
	}
	due to the monotonicity property of the logarithm.
	
	This path finding algorithm is deterministic on its own, however this would not account for real life scenarios. Two vehicles can take different routes between identical starting and ending locations (depending on random environmental factors such as traffic, weather, road blocks, etc.). Therefore, we introduce randomness into our trace generation by applying the \textit{Metropolis–Hastings algorithm} (MH) to the shortest path. Specifically, we have a target stationary distribution over all paths, where the probability of a path is computed as above from the routing graph. Sampling directly from this distribution is hard due to its finite but exponentially large domain, therefore, we rely on MCMC methods. Our MH algorithm applied to shortest paths is described in Algorithm \ref{alg:MH}. Markov chain theory says that we need multiple state transitions to have a ``good enough'' sample (that comes from a distribution close enough to the target), we perform \rev{10} transitions. We set this value based on our \textit{Route Distribution metric} (see Section \ref{sec:route_distro}) that measures the distance between original and synthetic routes taken between source-destination pairs. Our experiments in Section \ref{sec:experimental} show that $10$ iterations are sufficient for larger datasets, however, smaller databases benefit from $100$ and even $150$ iterations.
	
	
	The final step in our TG algorithm is \textit{looping}, where we aim to approximate the time a vehicle stays in one cell, that is, the number of repetitions of a single location in a trajectory. Looping allows to capture some traffic patterns  more faithfully such as rush hours or traffic jams. We model this by generating the repetition number $\eta$ of a location $L_x$ from a geometric distribution $\eta \sim Geom(1-TPG_{\theta_2}[L_x | t, L_x, L_{dst}])$, where $TPG_{\theta_2}[L_x | t, L_x, L_{dst}]$ is the probability output of TPG for staying at location $L_x$.

\begin{algorithm}[t]
	\small
	\caption{Metropolis-Hastings Algorithm for TG \label{alg:MH}}
	\DontPrintSemicolon
	\KwIn{Most probable path $p = (L_{src}=L_{\ell_1},L_{\ell_2},\dots L_{\ell_n}=L_{dst})$}
	Choose uniformly random node $L_{\ell_i} \sim U(p\setminus \{L_{\ell_1}, L_{\ell_n}\})$ \;
	Choose uniformly random neighbor $L_{\ell_i}^n$ of $L_{\ell_i}$\;
	Candidate path $p_c= (L_{\ell_1}, \ldots L_{\ell_{i-1}}, L_{\ell_i}^n, L_{\ell_{i+1}}, \ldots L_{\ell_n})$ \;
	Let $\gamma = \frac{\prod_{(L_x \to L_y) \in p_c}  TPG_{\theta_2}[L_{y} | t, L_{x}, L_{dst}]}{\prod_{(L_x \to L_y) \in p}  TPG_{\theta_2}[L_{y} | t, L_{x}, L_{dst}]}$\;
	$p = \begin{cases}
	p_c & \text{with probability} \min(1,\gamma) \\
	p & \text{otherwise}
	\end{cases}$\;
	\KwOut{$p$} 
\end{algorithm}

 
	
	\subsubsection{Remarks}
	Feeding the destination and time as an input to our transition probability generator enhances model accuracy by a large margin (in certain cases with more than 20\%). The rationale behind this is that the probability of the next-hop location is heavily influenced by the direction of movement, i.e., the specific destination where the individual is heading for. Similarly, time also impacts the direction of movement towards a specific destination, especially in vehicular transport, where the route of a vehicle is largely influenced by the traffic, i.e., ultimately time dependent. This is in sharp contrast to earlier works \cite{chen2012diffgrams} which solely used the last visited locations to predict the next location of a trajectory.
	

	\subsection{Privacy Analysis}\label{DP_params}
	In this section, we quantify the privacy guarantee of \our by using the moments accountant described in Section \ref{dp}.
    Recall that we use the Gaussian Mechanism to provide DP for the dimensionality reduction (Section \ref{sec:dimred}), and DP-SGD \cite{abadi2016deep} for TI (Section \ref{VAE}) and TPG (Section \ref{FFN}). 
    \our is the adaptive composition of these three mechanisms plus the trace generation (TG) described in Section \ref{TG}.
    \smallskip
    	
	\noindent \textbf{Dimensionality reduction:} To select the most frequent $K$ (top-$K$) cells, we employ the Gaussian mechanism (see Section \ref{dp}) and add i.i.d Gaussian noise $\mathcal{G}(L_{max} \sigma_K)$ to the visit counts of \emph{all} cells, where $L_{max}$ is an upper bound on the trace length, and return the cells as top-$K$ which have the $K$ largest noisy counts. \rev{Both $K$ and $L$ are chosen based on the threshold where 95\% of the \revs{visits} is included in our calculations.} 

	\smallskip
		
	\noindent \textbf{TI and TPG models:} To train the TI and the TPG models, we use the Differentially Private Stochastic Gradient Descent (DP-SGD) by Abadi et al. \cite{abadi2016deep}. This method is independent of the chosen loss function and model, and it adds noise to the clipped gradients. In particular, the gradients of all model parameters in every model update are clipped to have a bounded $L_2$-norm with value $C$, and then Gaussian noise with variance $C_{TI}^2 \sigma_{TI}^2$ (for TI) and $C_{TPG}^2 \sigma_{TPG}^2$ (for TPG) is added to the clipped gradients before updating the parameters. 
	The output of DP-SGD are the parameters $\theta_1$ and $\theta_2$ of TI and TPG, respectively. The sampling probability $q$ in DP-SGD is calculated as follows. 
	Our aim is to provide user-level (or in our case trajectory-level) Differential Privacy. However, recall that trajectories have different lengths, and we divided them into 1-grams: thus there are variable number of training examples belonging to a single trajectory for TPG. In the case of our TI model, we only have one sample per trajectory, thus the sampling probability of a single trajectory here is at most $|B|/|D|$, where $|D|$ is the total number of trajectories and $|B|$ is the size of batch $B$. However, for TPG, a single trajectory can have multiple samples, therefore, we sample batches differently. We first sample a trajectory from the dataset uniformly at random, and then a 2-gram out of this trajectory also uniformly at random. We repeat this experiment until a batch $B$ of grams (training samples for TPG) is collected. This sampling mechanism ensures that any trajectory is equally probable to participate in an update (batch), and hence the sampling probability becomes $q=\frac{|B|}{|D|}$.
	\smallskip
	
	\noindent \textbf{Trace Generation (TG):} The trace generation uses only the differentially private models $TI_{\theta_1}$ and $TPG_{\theta_2}$ as input, and does not access private data. Therefore, the generated traces are also differentially private due to the post-processing property of DP.
	\smallskip
	
	Let $\eta_0(x|\xi) =  \mathsf{pdf}_{\mathcal{G}(0, \xi)}(x)$ and $\eta_1(x|\xi) =  (1-q) \mathsf{pdf}_{\mathcal{G}(0, \xi)}(x) + q \mathsf{pdf}_{\mathcal{G}(1, \xi)}(x)$ where $q$ is the sampling probability of a single trace in a single round. Let
	\vspace{-.2cm}
	{\small
\begin{align}
\alpha_{K}(\lambda) &= (\lambda^2 + \lambda)/4\sigma_K^2 \label{eq:alphaK}\\
\alpha_{TI}(\lambda) &= \log\max(E_1(\lambda, \sigma_{TI}), E_2(\lambda, \sigma_{TI})) \\ 
\alpha_{TG}(\lambda) &= \log\max(E_1(\lambda, \sigma_{TPG}), E_2(\lambda, \sigma_{TPG})) 
\end{align}}
\vspace{-.2cm}
where
$
E_1(\lambda,  \xi) =  \int_{\mathbb{R}}\eta_0(x|\xi) \cdot \left(\frac{\eta_0(x|\xi)}{\eta_1(x|\xi)}\right)^{\lambda} dx
$,
$ E_2(\lambda,  \xi) = \int_{\mathbb{R}}\eta_1(x|\xi) \cdot \left(\frac{\eta_1(x|\xi)}{\eta_0(x|\xi)}\right)^{\lambda} dx
$, and $\alpha_{K}(\lambda)$, $\alpha_{TI}(\lambda)$, $\alpha_{TG}(\lambda)$ are the log moments (see Definition \ref{def:moments}) of the TOP-K identification, TI, and TPG, respectively. \rev{$\alpha_K(\lambda)$ in Eq.~\eqref{eq:alphaK} follows from Lemma 1 in \cite{acs2018diffmixture}}.

\begin{theorem}[\textbf{Privacy of \our}] \label{thm:our}
Let $e_{TI}$ and $e_{TG}$ denote the number of epochs for TI and TPG, and $|B|/|D|$ is the number of SGD iterations per epoch. \our is ($\epsilon$, $\delta$)-DP, where
{\small
\begin{align*}
\varepsilon = \min_{\lambda}\rev{\frac{1}{\lambda}}\left(\alpha_K(\rev{\lambda}) +  \frac{|B|}{|D|}e_{TI}\alpha_{TI}(\rev{\lambda})  + \frac{|B|}{|D|}e_{TG}\alpha_{\mathit{TG}}(\rev{\lambda}) - \log(\delta)\right) 
\end{align*}
}
\end{theorem}
\rev{\begin{proof} Let $\alpha_{DPLoc}(\lambda)(\lambda)$ denote the log moment DP-Loc in Alg.~\ref{alg:ours}. It follows from the second part of Theorem \ref{Moments} that $\varepsilon = \min_{\lambda}\frac{1}{\lambda}(\alpha_{DPLoc}(\lambda) - \log \delta)$. Since  dimensionality reduction is only applied once at the very beginning of DP-Loc, and there are $\frac{|B|}{|D|}e_{TI}$ SGD iterations in $TI$ and   $\frac{|B|}{|D|}e_{TG}$ SGD iterations in $TG$ in total, it also follows from the first part of Theorem \ref{Moments}   that $\alpha_{DPLoc}(\lambda) \leq \alpha_K(\lambda) +  \frac{|B|}{|D|}e_{TI}\alpha_{TI}(\lambda)  + \frac{|B|}{|D|}e_{TG}\alpha_{\mathit{TG}}(\lambda)$ which concludes the proof. 
\end{proof}}
Given a fixed value of $\delta$, $\varepsilon$ is computed numerically  as in \cite{abadi2016deep}.

	\section{Experimental evaluation}
	\label{sec:experimental}
	In this section, we empirically evaluate \our on \rev{three} publicly available datasets, where two contain the GPS trajectories of different taxi trips from San Francisco ~\cite{epfl-mobility-20090224} and Porto~\cite{porto}, \rev{and one has miscellaneous trajectories (e.g. walking, driving, public transport, cycling)} (mostly) from Beijing, China (\textit{GeoLife} dataset \cite{zheng2009mining, zheng2008understanding, zheng2010GeoLife}).
	We show that the synthetic trajectories generated by our model are close to the original trajectories according to \rev{four} different utility metrics, \revs{and apply a fifth utility metric to demonstrate the usefulness of the Metropolis-Hastings algorithm}.
	The architecture described in Section \ref{modeldesc} is fixed for all datasets. Although it has been shown that the optimization of hyperparameters (including the architecture of \our) is possible with DP guarantees~\cite{abadi2016deep,gupta2010dp}, yet this requires a larger privacy budget. For simplicity and owing to the similarity of the three datasets evaluated, we do not apply this technique here. 

	\subsection{Data}\label{Data}
	\revs{\textbf{San Francisco taxi: }}
	The original San Francisco (SF) taxi dataset contains a set of GPS trajectories with timestamps recorded by 536 taxis. They were collected over 30 days in the San Francisco Bay Area of the USA in 2009. The trajectories cover the region of San Francisco within the bounding box of (37.6017N, 122.5158W) and (37.8112N, 122.3527W) -- approximately $340\: \mathrm{km}^2$. The original sampling rate of these trajectories is roughly 1 per minute. We used all the $276 744$ trajectories from the SF dataset.\\
	\revs{\textbf{Porto taxi: }}
	The original Porto dataset contains 1.7 million GPS trajectories with timestamps of 441 taxis. This dataset was acquired in the metropolitan area of Porto, Portugal, over a period of nine months in 2012. The sampling rate varies, but it is approximately 1 per 15 seconds. We considered taxi trips only within a bounding box of (41.00456N, -8.7368W) and (41.2728N, -8.4412W) around the city, with a size of approximately $600\: \mathrm{km}^2$. We used a random sub-sample containing $450000$ trips for our evaluation.\\
	\revs{\textbf{GeoLife:}}
	\rev{The original GeoLife dataset is a collection of $17,621$ daily trajectories by 182 users in a period of over five years (from April 2007 to August 2012). $91.5\%$ of the trajectories are logged in a dense representation, e.g., every 1-5
seconds or every 5-10 meters per point. The GeoLife dataset includes a wide range of outdoor user movements, including commuting,  entertainment and sports activities, such as shopping, sightseeing, dining, hiking, and cycling. The majority of the data was generated in Beijing, China, where we set our bounding box of (39.75N, 116.2W) and (40.1N, 116.55W) of approximately $580\: \mathrm{km}^2$, and considered the traces within this region only. Note that Porto and GeoLife cover approximately the same area size, however, the Porto bounding box also contains its uptown area while GeoLife does not. In other words, the whole Beijing area is populated uniformly almost everywhere with approximately one seventh in available data points of the Porto dataset. Moreover, since the dataset contains continuous recordings of people's daily traces with multiple stationary locations, we cut these traces along their stationary locations and replace the original trace with its sub-traces in the dataset. More precisely, if an individual stayed at one location for a long time (more than 15 minutes), then the location is regarded as stationary and we terminated the given trace and initialized a new one with the current cell as new starting point. The justification behind such slicing is twofold. First, we intend to model and release only the mobility patterns of people which is the main focus of most practical applications (such as traffic optimization). Second, the original GeoLife dataset is small in terms of both individuals and traces; with the help of trace slicing we obtained $59,907$ traces. Note that (strictly speaking) DP-Loc provides DP guarantees only to individual sub-traces in this case; however, we embrace this limitation on two accounts: i) we could not obtain a larger suitable third dataset, ii) the limited number of users in the original GeoLife dataset would result in a low utility irrespective of the actual DP-based privacy-preserving mechanism applied.}
	
	\rev{Regarding the taxi datasets} our objective is  to preserve the privacy of passengers and not taxi drivers, and hence a trace (sequence) is composed of the recorded location visits of a single taxi trip.
	
	\subsection{Data Preprocessing}\label{preproc}
	We consider two grids with different cell size. The smaller grid consists of cells with size of $250 \times 250 \mathrm{m}^2$, and the larger one with cells of size $500 \times 500 \mathrm{m}^2$.\revs{ We have found that these two sizes are small enough to capture the movement of an individual, and also large enough to enable neural networks to learn the underlying distribution with sufficient accuracy and low computational cost \revss{(see Appendix \ref{app:complexity} the complexity analysis)}.
	Although, in a real-life scenario, a given application can change these sizes, we believe that most practical use-cases would not divert very much from these values.} 
	Each GPS location point is assigned to its covering cell. Therefore, every trace (taxi trip) is composed of the sequence of location visits, each containing a pair of cell and the time of the visit.
	All traces with velocity larger than $150$ km/h (calculated between two GPS points) or being out of the bounding box are dropped. 
	Since the sampling rate was not constant in the database, we applied two transformations to make it more regular; (1) cell visits are aggregated in time by $60$ seconds keeping the cell that was the most frequent in the trace during this time frame, and (2) when there were gaps shorter than 5 minutes without any location visits, these missing visits are approximated by linear interpolation. 
	We \revs{select the K cells that contain at least 95\% of all visits in a privacy-preserving manner} and we map each point to the closest top-K cell, if there is any within 1000m. We drop traces with locations that cannot be assigned to a top-K cell.
	\rev{Finally, if the resulting trace had only a single visit, it is dropped. All traces are truncated to length $L_{max}$.} \revs{Since DP adds noise to the cell counts, the values of $K$ and $L$ can change between runs; however, larger datasets are not prone to this instability. To illustrate this, we generated the values of $K$ and $L$ 5 times and included their average in Table \ref{tab:datasets}}. The standard deviation for Porto and SF were $0$, however, for GeoLife it was $20$ for the $K$ values; the $L$ values were stable in all cases. The data loss caused by top-K filtering was $3-5\%$ for all datasets. Less than $1\%$ of traces were dropped by the other filters.
	
	After cleaning and smoothing the data, the timestamps are further aggregated by assigning only the hour of the day, when the dominant part of the taxi trace was present, to all visits of a single trace. For example, if a trace started at 17:58, and ended at 18:10 with 12 visits altogether, we assigned the $18^{th}$ hour of the day to every cell visit of the trace (including those which happened before 18:00). As our aim is only to demonstrate the feasibility of our approach, this simplification is introduced in order to decrease the size of the input and output space making training faster and the models less complex.
	Finally, we created 2-grams from the traces, i.e., we grouped every two consecutive data points together to create a single training sample for our TPG model, where the first and second parts served as input and output for the model during training. The first part of every gram is augmented with the destination cell of the trace where the gram comes from, and the second gram contains only the cell identifier of the next location (without timestamp).

	Table \ref{tab:datasets} shows the descriptive statistics of the datasets. 
	
	\begin{table*}[t]
		\centering
		\small
		\begin{tabular}{|l|c|c|c|c|c|c|}
			\hline
			\textbf{Dataset} & $|D|$ & $topK$ & $\max |t|$ & $\mathrm{avg}|t|$ & $\mathrm{std.dev}|t|$ & $\mathrm{mode}|t|$\\
			\hline
			{\bf SF-250} & 431,222 & 550 & 19  & 10.54 & 6.67 & 7 \\ \hline
			{\bf SF-500} & 431,222 & 179 & 19 & 10.46 & 6.17 & 7 \\ \hline
			{\bf Porto-250} & 450,000 & 490 & 24 & 11.63 & 7.53 &9 \\ \hline
			{\bf Porto-500} & 450,000 & \rev{211} & 24 & 11.52 & 8.64 & 9\\ \hline
			{\bf \rev{GeoLife-250}} & 59,907 & 406 & 14 & 13.91 & 11.38 & 10 \\ \hline
			{\bf \rev{GeoLife-500}} & 59,907 & 503 & 22 & 13.86 & 11.38 & 10\\ \hline
		\end{tabular}
		\caption{The preprocessed datasets used in our experiments: SF-250, Porto-250, \rev{GeoLife-250} (with a cell size of 250$m^2$) and SF-500, Porto-500, \rev{GeoLife-500} (with a cell size of 500$m^2$). Trace length $|t|$ is for traces $t \in D$ .}
		\label{tab:datasets}
	\end{table*}

	\subsection{Evaluation Metrics}\label{metrics}
	Choosing a truly representative metric is always challenging and, to a large extent, application-dependent. We focused on some commonly used characteristics of PoIs, heatmaps, popular start and destination places, rush hours, traffic jams, trip lengths, etc. 
	As a result, we consider four different utility metrics \rev{which are mainly borrowed from previous works \cite{chen2012diffgrams, gursoy2018utility}}. Each of them is evaluated both on the synthetic and the original dataset, and the difference is measured according to different distance metrics.

	\subsubsection{Trip size distribution}
	The Jensen-Shannon divergence (JSD) is computed between the distribution of the trip lengths in the synthetic and the original datasets in each hour of the day (the trip length is the number of cells of a trip). 
	Note that, unlike the Kullback-Leibler (KL) divergence, JSD is symmetric and has a finite value. In our case, JSD is bounded between 0 (identical distributions) and 1 (least similar distributions).
	
	\subsubsection{Frequent patterns}
	The top-$N$ most frequent patterns (i.e., subsequences of locations) are computed both in the original $D$ and synthetic dataset $D'$, which are denoted by $F_N(D)$ and $F_N(D')$, respectively.  The true positive ratio $\frac{|F_N(D) \cap  F_N(D')|}{N}$ is reported for $N=10, 20, 50, 100$.
	
	\subsubsection{Spatio-temporal distribution of location visits}
	The spatio-temporal density in a given hour of the day is the number of visits in each cell of the considered region. This histogram, where each bin corresponds to a cell, is computed from both the synthetic and the original traces individually, and the spatio-temporal density of the original dataset as well as the synthetic datasets are obtained from these histograms after normalization. The Earth Mover's Distance~\cite{emd} is reported between these distributions, which measures their difference in terms of geographical distance (meters) and is a metric for probability distributions. Specifically, EMD measures the ``amount of energy'' (or cost) needed to transform one distribution to another where the ground distance is the geographical distance between the centers of cells. The EMD between the spatial densities of the original and synthetic data are reported for cells that include at least $80\%$ of the data, but no more than 2000 cells for every hour, and also over all hours of the day.
	
	\subsubsection{Spatio-temporal distribution of source and destination pairs} The joint distribution of the source and destination locations is computed from the original and synthetic datasets individually, and their EMD is reported for every hour, and over all hours of the day. In particular, we count the relative frequency of every possible pair of source and destination locations in both datasets, and compute the EMD between these two distributions as above (the distance between a pair of location points is the sum of their individual distances). As the domain of this joint distribution has a size of $|\mathbf{L}|\times |\mathbf{L}|$ \revs{(in every hour, not over all hours)}, the computation of EMD can be very costly. We used the same approximation as for EMD density described above. \revs{Note that besides reporting the hourly EMD, we also calculated the EMD averaged over all hours in order to compare with prior works.}
	
	\subsubsection{Route distribution}
	\label{sec:route_distro}
	 \rev{We calculate the histograms of cell visits for the inner points of a trace conditioned on the source and destination pairs, i.e. the histogram shows the distribution of cells that were visited between a given source and destination. The route distribution is calculated for both original and synthetic datasets, then their EMD is measured for each source-destination. Due to the high number of such pairs, we report the average of their EMD values. Intuitively, this metric measures how ``realistic'' a synthetic trajectory is. Due to the very high number of source and destination pairs we used the same approximation described above.} 
	 
	 	\vspace{-0.2cm}
    \subsection{Baselines}
    
     As comparison to \our, we evaluate the Ngram model from \cite{chen2012diffgrams} (Ngram), and the AdaTrace model from \cite{gursoy2018utility}. We chose AdaTrace because it is the state-of-the art technique and outperforms several other models (Ngram, DPT, SGLT) as stated in \cite{gursoy2018utility}. Ngram is chosen because it performs very well regarding frequent patterns and is a more general solution for any sequential data (unlike AdaTrace). For the reasons explained in Section \ref{sec:related}, we do not compare \our to SGLT \cite{SGLT}, since it is incompatible with the mobility patterns represented by our datasets.  Moreover, our goal is to generate large synthetic datasets, and SGLT was shown \cite{gursoy2018utility} not to scale well in this regard\footnote{We also tried to run DPT, but even for 100.000 traces it ran out of 100GB of memory.}.
   
    Since AdaTrace uses a dynamic grid configuration we had to align our grid in order to reach a fair comparison. First, we transformed the GPS coordinates with Mercator projection and applied all our preprocessing steps to the records (even top-$K$ selection). Next, we generated synthetic traces with AdaTrace, where the output is also in the Cartesian space. For the evaluation with our metrics, we assigned the output coordinates to our non-dynamic grid cells. 
    
	As AdaTrace and Ngram apply the Laplace mechanism, for fair comparison, we changed it for the Gaussian mechanism and calculated the values of $\epsilon$ by the moments accountant method. For both models, given the value of $\delta=1/|D|$ and \rev{$\epsilon=[0.5, 1,2,5]$}, we analytically calculated the variance of the noise using Theorem \ref{Moments}. The $L_2$ sensitivity of the Adatrace can be shown to be $\Delta_2 = 1$ based on \cite{gursoy2018utility}.
	As for Ngram, we bounded the $L_2$ sensitivity with the $L_1$ for the same reason as in our top-$K$ selection (see Section \ref{DP_params} for details).
	\vspace{-0.2cm}
	
	\subsection{Experiment setup}
	
\begin{figure*}
        \begin{center}
        \begin{subfigure}[]{0.3\textwidth}
                \includegraphics[width=\textwidth]{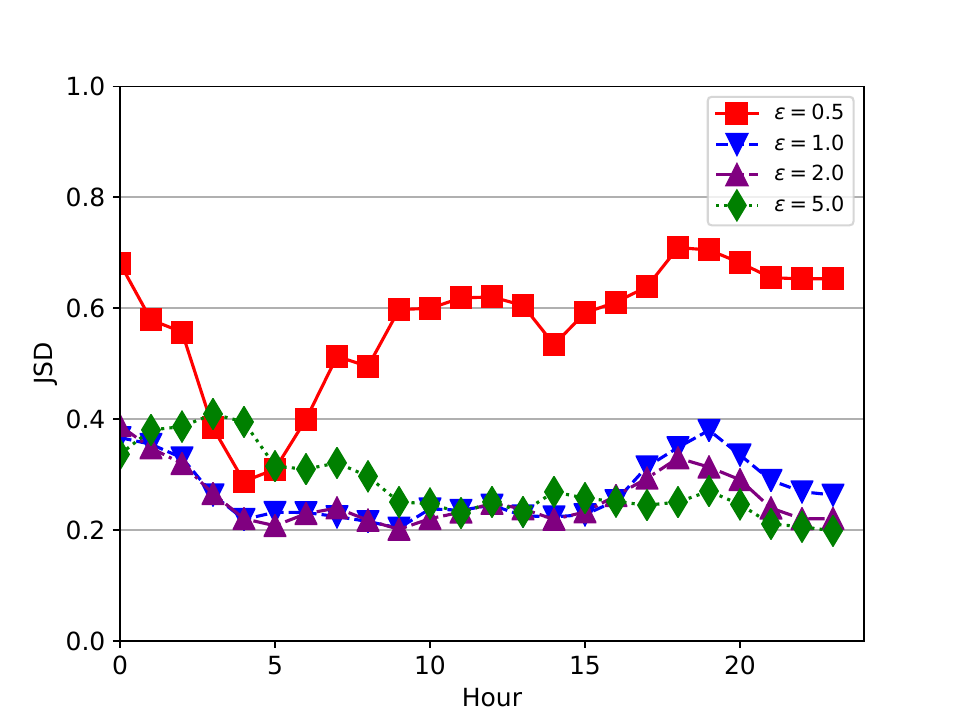}
                \caption{JSD, cell size: 250 m}
                \label{fig:jsd250porto}
        \end{subfigure}
          \vspace{-0.08cm}
        \begin{subfigure}[]{0.3\textwidth}
                \includegraphics[width=\textwidth]{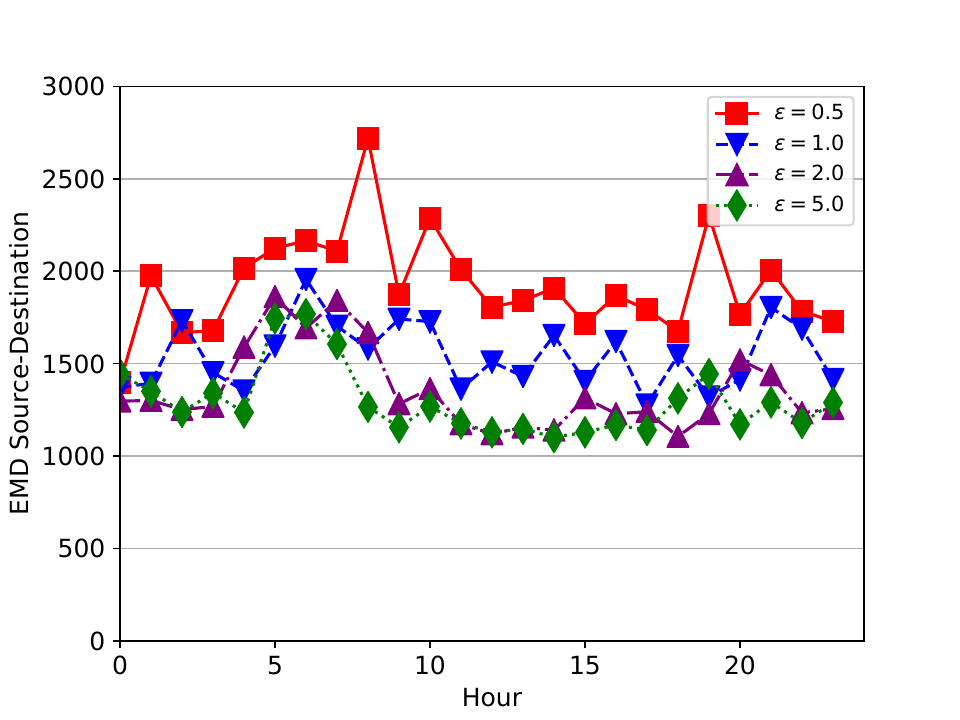}
                \caption{EMD-SD (VAE), cell size: 250 m}
                \label{fig:emdvae250porto}
        \end{subfigure}
          \begin{subfigure}[]{0.3\textwidth}
                \includegraphics[width=\textwidth]{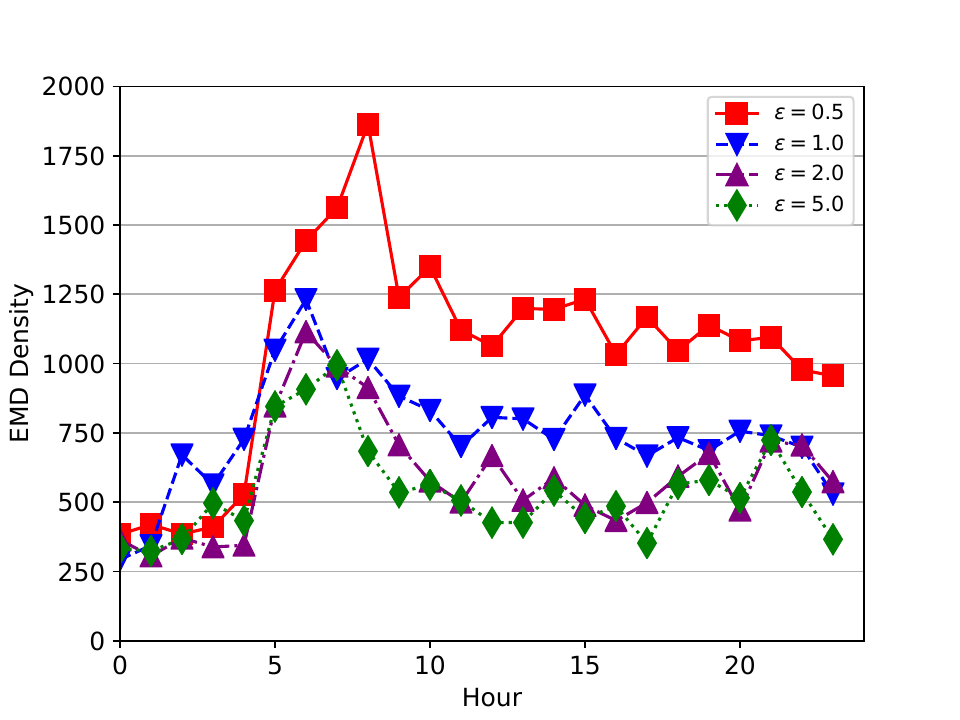}
                \caption{EMD-Density, cell size: 250 m}
                \label{fig:emddens250porto}
        \end{subfigure}
         \begin{subfigure}[]{0.3\textwidth}
                \includegraphics[width=\textwidth]{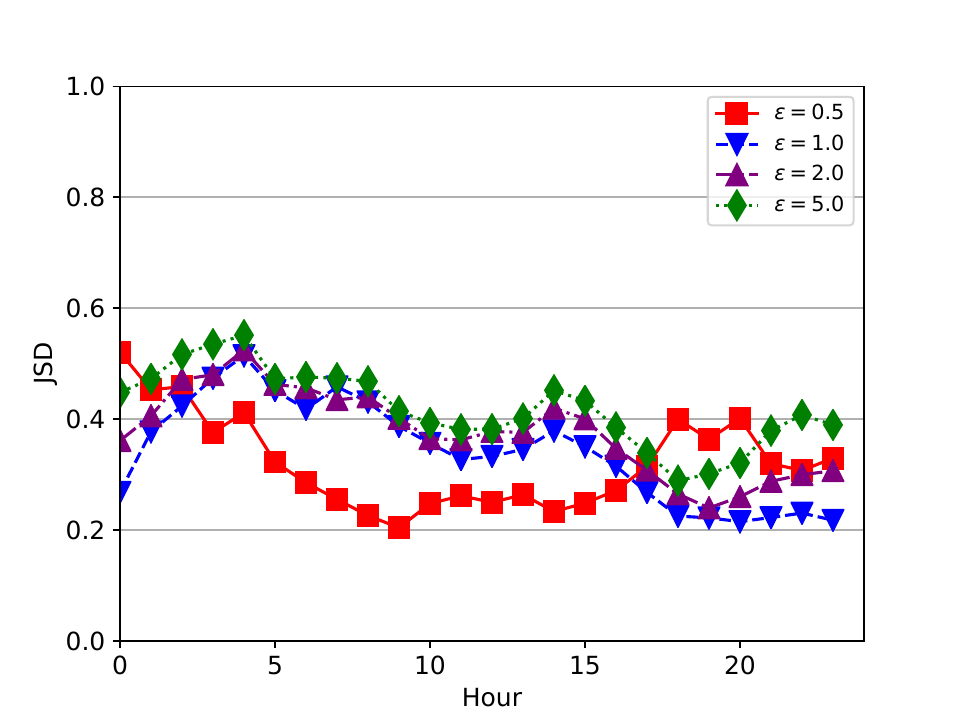}
                \caption{JSD, cell size: 500 $m$}
                \label{fig:jsd500porto}
        \end{subfigure}
        \begin{subfigure}[]{0.3\textwidth}
                \includegraphics[width=\textwidth]{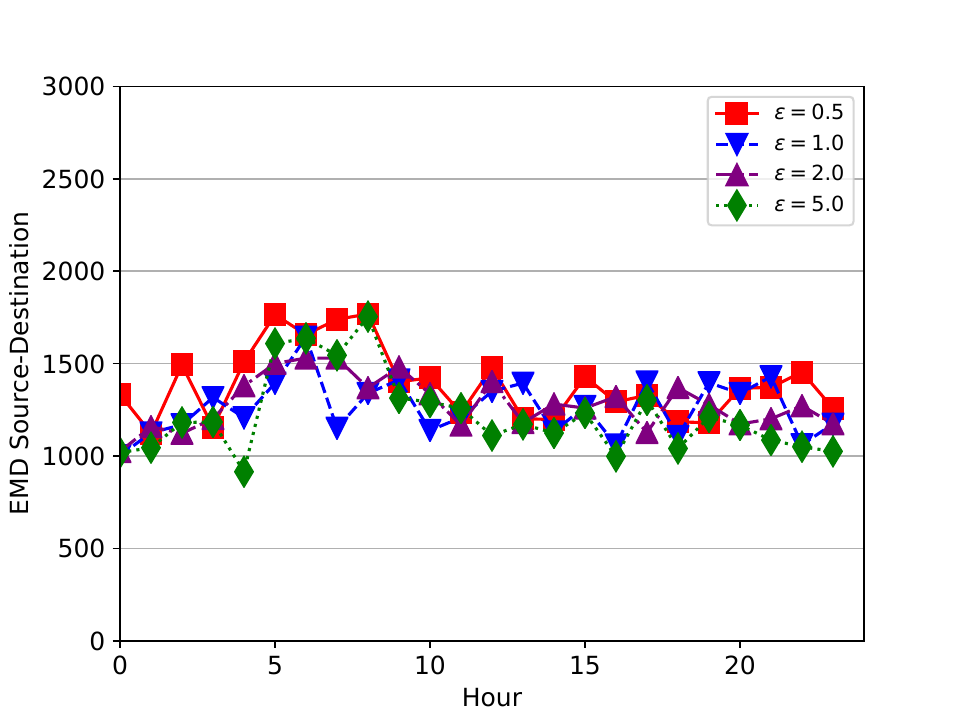}
                \caption{EMD-SD (VAE), cell size: 500 $m$}
                \label{fig:emdvae500porto}
        \end{subfigure}
    \begin{subfigure}[]{0.3\textwidth}
                \includegraphics[width=\textwidth]{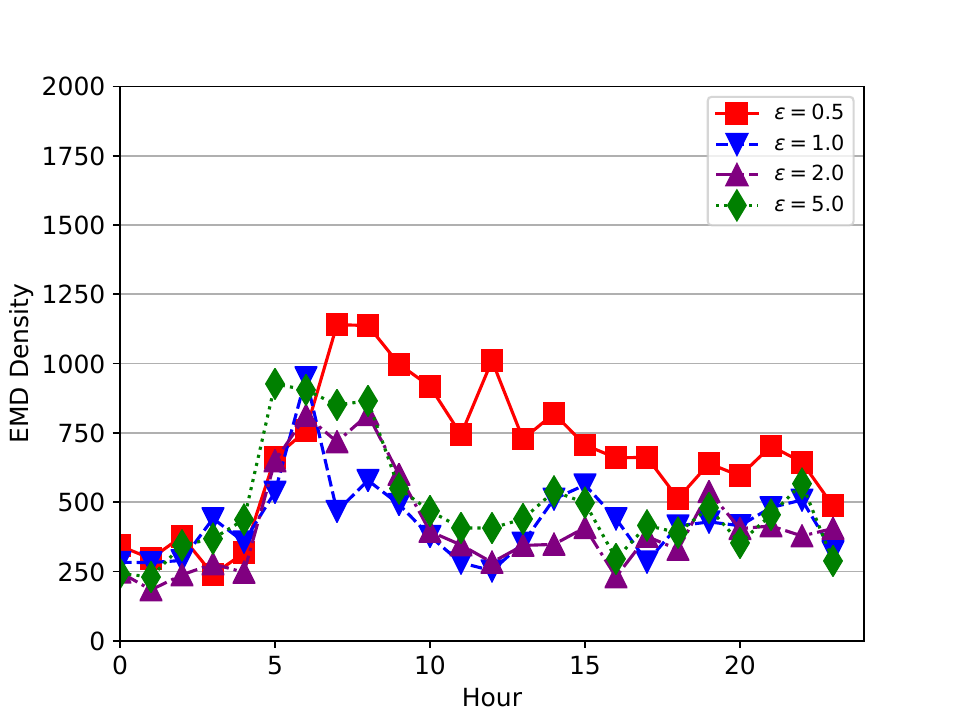}
                \caption{EMD-Density, cell size: 500 $m$}
                \label{fig:emddens500porto}
        \end{subfigure}
        \vspace{-0.2cm}
        \caption{Performance of our approach on Porto dataset depending on the time ($\delta=4\cdot10^{-6}$).}
    \label{fig:perf_on_time}
    \vspace{-2mm}
    \end{center}
\end{figure*}

		We experimentally evaluate the performance of our solution in terms of the above four utility metrics on the datasets described in Table \ref{tab:datasets}. We also present the results of Ngram and AdaTrace on the same datasets, and compare these to our work using the four metrics.
	As neither Ngram nor AdaTrace releases time information, we dropped all timestamps in all datasets, and synthesized the resulting datasets with the two models \revs{(by contrast, \our was always executed on the original datasets with temporal information, and its results are reported in Figure \ref{fig:perf_on_time} and \ref{fig:perf_on_time_sf}}). We obtained the implementation from the respective authors.
	Experiments were conducted using Tensorflow 2.0 and Python 3.6.9 on a single Linux server with 98GB RAM and 16 cores. 
	
		We have carried out our \rev{training in 24 different settings, combining $\epsilon = 0.5,1,2,5$  with two grid resolutions (250 and 500), and three datasets (for statistics see Table \ref{tab:datasets}).} 
	For the TI model, we set the $L_2$-norm clipping threshold $C_{TI}$ to $1.00$, the batch size $|B|$ to $200$, and ran the training over $e_{TI}=15$ epochs (recall that  $C_{TI}^2 \sigma_{TI}^2$ is the variance of the Gaussian noise added to the gradients in order to provide DP). With $\epsilon=1$ and $\sigma_{TI}=1.5$, the learning rate was set to $0.2$; with $\epsilon=2$ and $\sigma_{TI}=1.00$, the learning rate was set to $0.5$; and with $\epsilon=5$ and $\sigma_{TI}=0.7$, the learning rate was $0.5$. Unless otherwise noted, we use $\delta=1/|D|$ in the rest of the paper, \rev{and $\epsilon$ is the overall privacy budget of DP-Loc}.
	For the TPG model, we set the $L_2$-norm clipping threshold $C_{TPG}=3.0$, the batch size to $200$, and trained the model over $e_{TPG}=15$ epochs. With $\epsilon=1$ and $\sigma_{TPG}=1.6$, the learning rate was set to $0.1$; with $\epsilon=2$ and $\sigma_{TPG}=1.00$, the learning rate was $0.15$; and with $\epsilon=5$ and $\sigma_{TPG}=0.6$, the learning rate was $0.15$.
	In dimensionality reduction, we add Gaussian noise to the counts of all location visits (when choosing the top-$K$ locations, see Section \ref{DP_params}) with $\epsilon=1$, $\sigma_K=3.8$; $\epsilon=2$, $\sigma_K=1.9$; and $\epsilon=5$, $\sigma_K=1.6$. 
	The total $\epsilon$ of \our is computed according to Theorem \ref{thm:our} over $15$ epochs for each of the TI and TPG models ($e_{TI} = e_{TPG}= 15$).
		
   \vspace{-0.2cm}
	\subsection{Results}\label{results}	
	In Figure \ref{fig:perf_on_time} and \ref{fig:perf_on_time_sf}, we report the hourly JSD and two EMD values between the original and the synthetic data generated by \our for $\epsilon=[0.5,1,2,5]$. Figure \ref{fig:perf_on_time} (and \ref{fig:perf_on_time_sf} in Appendix) show how the granularity of the grid influences the impact of the noise. Comparing the first and second lines of Figure \ref{fig:perf_on_time} and \ref{fig:perf_on_time_sf}, one can see that a larger grid size results in less error, though with coarsened data. (Coarsening the granularity eventually results in zero error of EMD, since all points end up in the same cell.)  For comparison, we report the overall JSD, EMD-SD, EMD-Density and Frequent Patterns results of our synthetic traces and that of Ngram and AdaTrace in Table \ref{table:results}. \rev{GeoLife follows the same trends, thus we included the more comparative results in Table \ref{table:results}}. \smallskip
	
	\noindent \textbf{Trip sizes}: \revss{In Figure \ref{fig:perf_on_time} and \ref{fig:perf_on_time_sf}, JSD shows the same trend and takes up very similar values for $\epsilon = 1,2,5$. However, this is not true for $\epsilon = 0.5$, since, to ensure such a low $\epsilon$ value, the algorithm requires considerably more noise. The San Francisco dataset shows the same tendency for JSD values (see Appendix \ref{app:complexity} for additional results). In the $\epsilon = 0.5$ case, JSD values are much larger when the cell size is 250m, also we can see that the EMD-SD values are also high. The TI model produced cells that were too close to each other, that caused very short traces ($4.8$ in average) in return.}
	Table \ref{table:results} shows that a larger grid generally results in larger JSD, but smaller EMD values for \our and AdaTrace, but larger for Ngram.
  	\emph{Recall that Ngram and AdaTrace do not include time information}, thus we only report one value for each setting in Table \ref{table:results}. To ease readability, we colored the best values at each metric among the three models in red.
    \our's JSD results are clearly much lower (i.e., closer to the original distribution) than that of the other two models. \rev{The MH iterations and looping extension play a large part in \revs{these} low results, but also by leaving out these steps we still get much lower JSD values than Ngram and AdaTrace (experiments show that such a JSD results in approximately a value of 0.5 JSD)}. Ngram generates traces without destination, and it does not select the globally most likely trajectory. In contrast to this, \our is more realistic. Although AdaTrace does include the destination, it still performs much worse than \our regarding trip sizes. We hypothesize that the high JSD values for AdaTrace could be due to the dynamic construction of the grid. AdaTrace works in two layers only and it is probable that many areas are not optimally divided, and our uniform grid with top-K selection performs better. Moreover, for Ngram and AdaTrace, the average generated trace length was approximately 3 (in all settings); both the mode and standard deviation were 2. In contrast to these, \our generates traces with an average length of 12, mode of 8 and standard deviation of 5, which are almost identical to the original statistics in Table \ref{tab:datasets}. 
    \begin{figure*}
        \centering
    \begin{subfigure}[]{0.24\textwidth}
                \includegraphics[width=\textwidth]{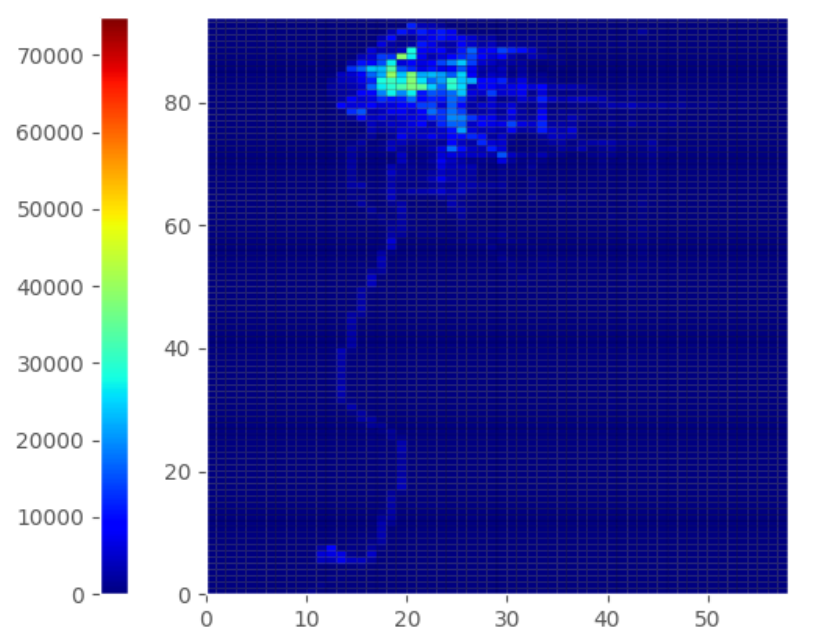}
                \caption{Original,cell size: 250 m}
                \label{fig:heatmaporig}
        \end{subfigure}
          \vspace{-0.08cm}
        \begin{subfigure}[]{0.24\textwidth}
                \includegraphics[width=\textwidth]{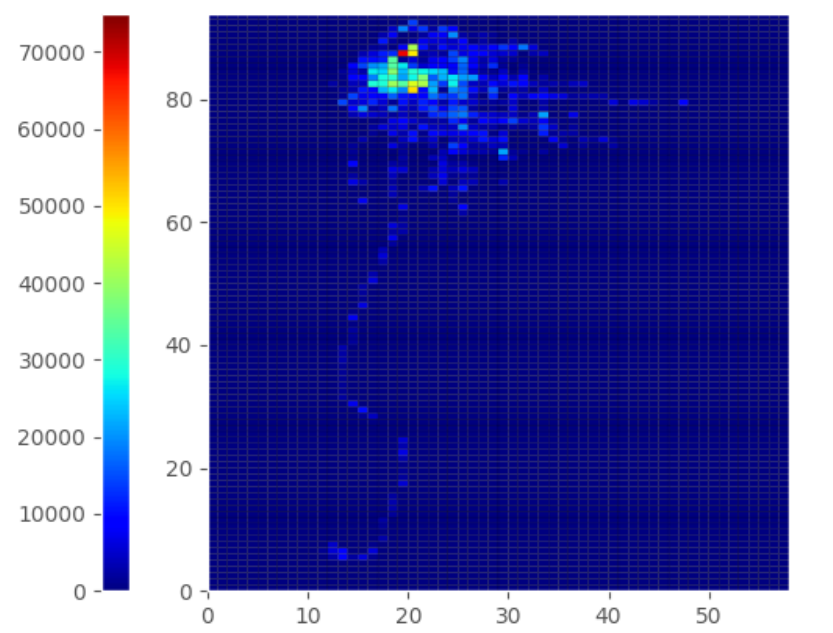}
                \caption{\our, cell size: 250 m}
                \label{fig:heatmapour}
        \end{subfigure}
          \begin{subfigure}[]{0.24\textwidth}
                \includegraphics[width=\textwidth]{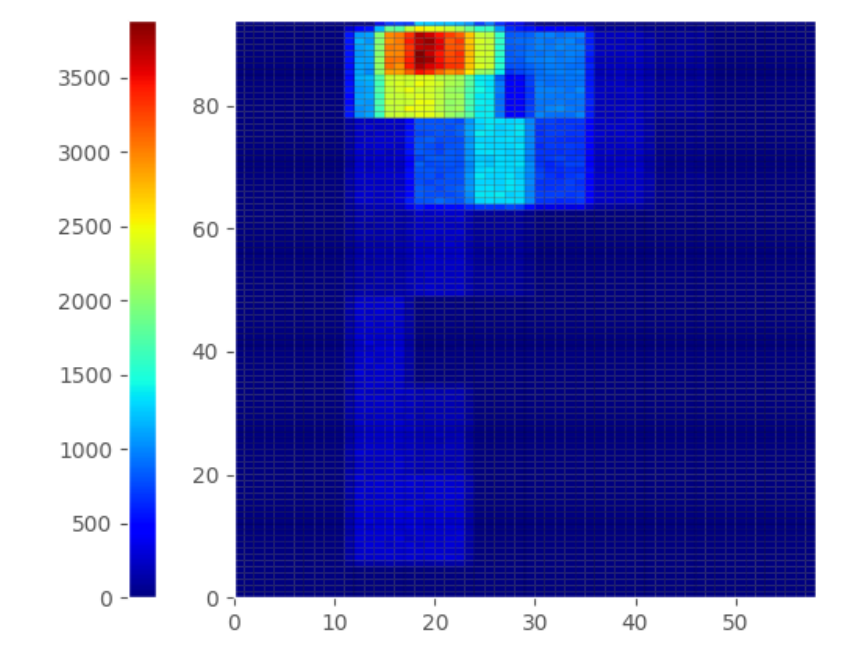}
                \caption{AdaTrace, cell size: 250 m}
                \label{fig:heatmapada}
        \end{subfigure}
         \begin{subfigure}[]{0.24\textwidth}
                \includegraphics[width=\textwidth]{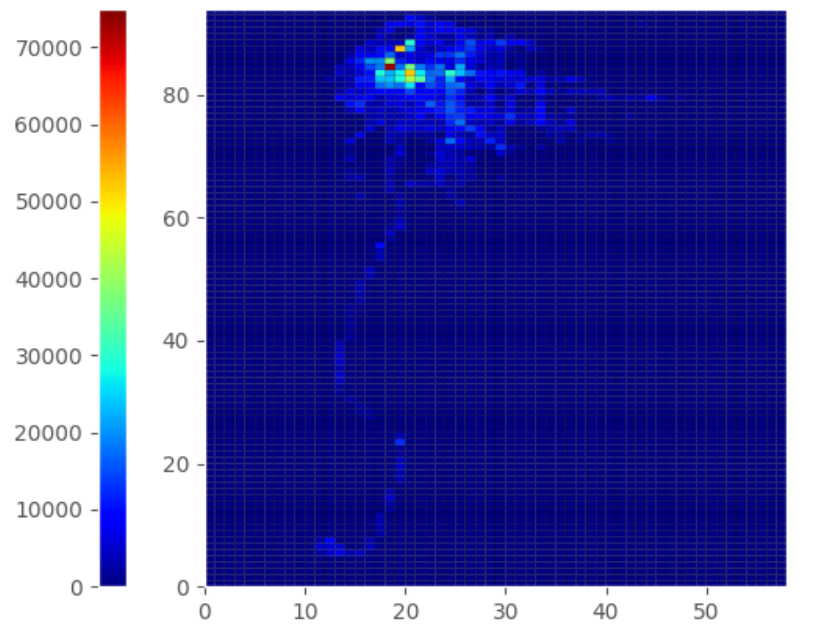}
                \caption{Ngram, cell size: 250 m}
                \label{fig:heatmapngram}
        \end{subfigure}
        \vspace{-0.2cm}
        \caption{Heatmaps of the synthetic and original databases.}
    \label{fig:heatmaps}
    \end{figure*}
    		
	\begin{table*}[h]
        \caption{Result of our DP-Loc algorithm without Differential Privacy \revs{and with 10 MH iterations}} \label{table:nodp}
        \centering
        \begin{small}
	\begin{tabular}{l|c|c|c|c|c|c|c|c|c}
	\hline
			\textbf{Dataset} & FP-10 & FP-20 & FP-50 & FP-100 & JSD len & avg len & std. dev. len & EMD src-dst & EMD density\\
			\hline
			{\bf SF-250} & 0.54 & 0.60 & 0.78 & 0.78 & 0.362 & 13 & 5.6 & 1572 & 602 \\ \hline
			{\bf SF-500} & 0.80 & 0.80 & 0.88 & 0.88 & 0.322 & 14 & 6.1 & 1669 & 629 \\ \hline
			{\bf Porto-250} & 0.70 & 0.90 & 0.90 & 0.92 & 0.296 & 13 & 5.3 & 1113 & 345 \\ \hline
			{\bf Porto-500} & 0.77 & 0.80 & 0.85 & 0.87 & 0.360 & 15 & 6.3 & 1201 & 367\\ \hline
			\rev{{\bf GeoLife-250}} & 0.90 & 0.90 & 0.90 & 0.90 & 0.408 & 10 & 3 & 4211 & 1201 \\ \hline
			\rev{{\bf GeoLife-500}} & 1.00 & 1.00 & 1.00 & 1.00 & 0.480 & 10 & 3.5 & 4746 & 1116\\ \hline

	\end{tabular}
	\end{small}
\end{table*}
   Ngram represents the dataset with a prefix tree which contains the set of all grams occurring in the dataset along with their occurrence counts, i.e. Ngram adds noise to these counts and prunes the noisy tree by removing grams with too small noisy count in order to improve accuracy.
    In particular, Ngram keeps grams only with sufficiently large occurrence counts, and shorter grams have larger counts which tend to survive  sanitization unlike longer grams \cite{chen2012diffgrams} with generally smaller counts. As a result, stronger privacy requirement (smaller $\epsilon$) results in a set of shorter grams and eventually a smaller sanitized prefix tree. Although shorter sanitized grams can be combined into longer grams in Ngram, the number of possible extensions was not high in our datasets. The finally released grams coincide with the set of most frequently visited places and hence Ngram preserves the statistics of most frequent patterns accurately. 
    The same holds for AdaTrace as well, because it also follows the Markov assumption to generate traces.

		\begin{table*}[!htb]
        \caption{Result measured on a downtown area in Porto. \revs{$\epsilon =1$, and 10 MH iterations} } \label{table:downtown}
        \centering
        \begin{small}
	\begin{tabular}{l|c|c|c|c|c|c|c}
	\hline
			\textbf{Dataset} & FP-10 & FP-20 & FP-50 & FP-100 & JSD len & EMD-src-dst & EMD-Density \\ \hline
			{\bf Ngram} & 0.90 & 0.90 & 0.84 & 0.98 & 0.878 & 1766 & 74 \\ \hline
			{\bf AdaTrace} & 0.40 & 0.50 & 0.64 & 0.58 & 0.709 & 1488 & 336 \\ \hline
			{\bf \our} & 0.90 & 0.85 & 0.95 & 0.90 & 0.278 & 1018 &  311 \\ \hline
	\end{tabular}
	\end{small}
\end{table*}
	\begin{table*}[!htb]
	\caption{Results of route EMD for different MH iterations for $\epsilon = 1$ (measured in meters)} \label{table:MH}
        \centering
        \begin{small}
	\begin{tabular}{l|c|c|c|c|c|c|c|c}
	\hline
		 & 0 & 1 & 5 & 10 & 25 & 50 & 100 &150 \\
			\hline
		{\bf SF-500} & 1335 & 1252 & 1177 & \red{1154} & 1171 & 1177 & 1180 & 1179 \\ \hline
		{\bf Porto-500} & 1347 & 1229 & 1137 & \red{1116} & 1119 & 1126 & 1134 & 1135  \\ \hline
    	{\bf GeoLife-500} & 1907 & 1693 & 1918 & 1794 & 1361 & 1150 & 1121 & \red{1119} \\ \hline
	\end{tabular}
	\end{small}
\end{table*}

\noindent \textbf{EMD-SD}: In Figure \ref{fig:emdvae250porto}, \ref{fig:emdvae500porto}, \ref{fig:emdvae250sf} and \ref{fig:emdvae500sf}, EMD is reported between the spatial distribution of the source and destination pairs depending on the time of the day. In Figure \ref{fig:emdvae250porto} and \ref{fig:emdvae500porto}, \revss{EMD values stay steadily between 900 and 2000 meters except in Figure \ref{fig:emdvae250porto}, where the trend for $\epsilon = 0.5 $ is similar but the values exceed 2500m. The results for the four different values of $\epsilon$ are almost the same if the larger cell size is used, and keep a steady distance in the 250m case.}
	Table \ref{table:results} shows that \our outperforms Ngram and AdaTrace, and generates more realistic endpoints for the traces. This demonstrates the generative power of the applied VAE network. The San Francisco dataset shows slightly different results. \revss{The EMD values in Figure \ref{fig:perf_on_time_sf} are almost the same for different values of $\epsilon$, except for $\epsilon = 0.5$ in Figure \ref{fig:emdvae250sf} where EMD values are higher}. We can also see that around 2 pm there is a peak in the two EMD values. We hypothesize that this is due to insufficient quality and quantity of data at that time period, which is not present in the Porto dataset. Nonetheless, \our still gives outstanding result in most cases. \revs{Note that smaller values are better, and EMD sums up all costs over the whole distribution. The distance between two neighboring cells is 250/500 m (center-to-center); if, e.g., EMD equals to $1000$m, it is the size of 2/4 cells over the whole distribution}.
	
	\noindent \textbf{EMD-Density}: In Figure \ref{fig:emddens250porto}, \ref{fig:emddens500porto}, \ref{fig:emddens250sf} and \ref{fig:emddens500sf}, EMD is reported between the distribution of location visits for every hour of the day. \revss{Values show a trend similar to EMD-SD, but remain below $1000$m in most cases, and $\approx 6-700$m on average. The values for $\epsilon = 0.5$ again keep a distance from the other results with a peak just below 2000m for the smaller cell size.} Ngram outperforms \our and Adatrace (except on Porto-500 where \our is the best). However, Ngram was originally designed to focus on the accurate release of the most frequent subsequences, and reconstructs traces from these. Therefore, as long as the number of visits per cell follows a power-law distribution, Ngram is expected to remain superior. Nevertheless, Table \ref{table:results} also shows that \our significantly outperforms AdaTrace in all cases. Figure \ref{fig:heatmaps} shows the San Francisco heatmaps of the synthetic datasets generated by all three models compared to the original one in Figure \ref{fig:heatmaporig}. Note that the scale of AdaTrace is lower than that of the rest (using the same scale would render AdaTrace's patterns almost invisible). We can see that the original (left) image is more detailed than any other image, but \our and Ngram imitate it closely. The original image contains less highly populated areas (deep red), and Ngram has the densest cell of all (ruby red) in downtown SF. 

	\noindent \textbf{Frequent Patterns:} Table \ref{table:results} reports the overall results of the \textit{Frequent Patterns} metric, that is, the true positive ratio of the top-$N$ location subsets between the original and the private synthetic databases. For all datasets, \our and Ngram perform similarly; from all \rev{120} cases \our outperforms Ngram \revss{$92$} times, and underperforms \revss{$73$} times, but in most cases the margin is very small. In all cases the performance of \our is higher than that of AdaTrace. The low FP values of \our regarding the GeoLife dataset \revs{show} that the dataset size was insufficient for the neural networks to learn the underlying distribution. Beijing is approximately twice as big as Porto, however, the data available in the GeoLife dataset was only 1/8th of that.
	Comparing Figure \ref{fig:heatmaps} along with the FP results to the JSD results, we can see why there is not a single universal metric for fair comparison. Ngram shows very good results when it comes to density (and frequent patterns), but the generated traces are highly unrealistic. \our might not perform as outstanding as Ngram regarding density metrics, but it produces realistic timestamped trajectories.
	
	\noindent \rev{\textbf{\our on smaller bounding-box:} The San Francisco and Porto datasets contain the whole urban area with their airports included on the map. For fair comparison we have run \our on the downtown area of Porto, similarly to \cite{gursoy2018utility}. The results are shown in Table \ref{table:downtown}; we only evaluated the downtown area in one scenario where the cell size is $500m$ and $\epsilon = 1$. It is visible that the performance of AdaTrace is better than on a larger grid, however, the ratio among the best results is the same as measured with a larger bounding-box. }

\noindent \textbf{\our without DP:} We also evaluated \our without Differential Privacy, i.e. no noise was added to the model at any stage. Table \ref{table:nodp} shows the results for the non-privacy preserving generated datasets. We can see that the numbers are almost identical to the ones in Table \ref{table:results}. The JSD and FP values are similar, and the two EMD values are lower than their private counterpart.

	\noindent \rev{\textbf{Justification for MH:} We applied the Metropolis-Hastings algorithm to the shortest paths in order to obtain convergence to a target stationary distribution over all paths, where the probability of a path is computed from the routing graph. In Table \ref{table:MH} we report the results of the route EMD metric, i.e., the EMD distance (in meters) between the original and the synthetic routes (cells) taken between source and destination pairs. Results show that for large datasets 10 iterations of MH results in the most realistic traces. However, for smaller dataset, where the added noise \revs{has} a higher influence on the generated traces, 100 and 150 iterations of MH show sufficient improvement. Consequently, after 10 or 100 iterations we could also observe a small drop in the EMD-Density values as well.}

\vspace{-0.56cm}
\section{Conclusions}
	\label{sec:conclusion}
	We proposed a novel approach to release location data with strong privacy guarantees. In contrast to prior works, \our is capable to release time information along with location visits without suffering significant utility loss. Our framework consists of generating the source and destination pairs of every trace, computing the transition probabilities between neighboring locations, then generating synthetic trajectories between the source and destination using a Monte Carlo algorithm. The transition probability depends on the time and destination, and is computed between the most frequent locations. We evaluated our proposal on three public location datasets and designed neural networks to model the distribution of trajectories. These networks are simple and hence fast to train even with DP guarantees. Results show that the provided utility is meaningful. Therefore, our technique can be a compelling new approach to the privacy-preserving release of complete location trajectories with time information. Importantly, we produce synthetic datasets that preserve many different statistics of the original dataset. Undoubtedly, releasing only a few targeted statistics with or without Differential Privacy, instead of the complete synthesized dataset, is a different approach which should always result in greater accuracy but only with respect to the released statistics. The proposed framework is general and finding the best generative models to a given type of data is difficult and requires domain expertise. We believe that our general approach may be applicable to other types of sequential data than location trajectories such as different time series.
	
\begin{landscape}
	\begin{table}[htbp]
        \caption{Summary of results \revs{with 10 MH iterations}. AdaTrace and Ngram ignores the time of trips, hence we report the the overall JSD and EMD values over the whole period for our approach. Best values are in red.} \label{table:results}
        \centering
        \resizebox{1\columnwidth}{!}{%
        \begin{small}
	\resizebox{\columnwidth}{!}{\begin{tabular}{lr|lll|lll|lll|lll|lll|lll}
	\cline{3-20}
		\multicolumn{2}{l|}{} & \multicolumn{6}{c|}{SF} & \multicolumn{6}{c|}{Porto}& \multicolumn{6}{c|}{GeoLife}\\
		\cline{3-20}
		\multicolumn{2}{l|}{} & \multicolumn{3}{c}{250} & \multicolumn{3}{c|}{500} & \multicolumn{3}{c}{250} & \multicolumn{3}{c|}{500}& \multicolumn{3}{c}{250} & \multicolumn{3}{c|}{500}\\
		\cline{3-20}
		\multicolumn{2}{l|}{} & \multicolumn{1}{l}{Ngram} & \multicolumn{1}{l}{Ada} & \multicolumn{1}{l}{\our} & \multicolumn{1}{l}{Ngram} & \multicolumn{1}{l}{Ada} & \multicolumn{1}{l|}{\our} & \multicolumn{1}{l}{Ngram} & \multicolumn{1}{l}{Ada} & \multicolumn{1}{l}{\our}  & \multicolumn{1}{l}{Ngram} & \multicolumn{1}{l}{Ada} & \multicolumn{1}{l|}{\our}& \multicolumn{1}{l}{Ngram} & \multicolumn{1}{l}{Ada} & \multicolumn{1}{l}{\our}  & \multicolumn{1}{l}{Ngram} & \multicolumn{1}{l}{Ada} & \multicolumn{1}{l|}{\our}\\
		\hline
		\hline

		& $\epsilon=0.5$ &\red{0.80}&0.20&\red{0.80}&\red{0.90}&0.20&0.80&\red{0.80}&0.20&0.70&\red{0.90}&0.28&0.60& \red{0.66}&0.05 & 0.30&\red{0.70}&0.01&\red{0.70}\\
		& $\epsilon=1$ &\red{0.80}&0.10&\red{0.80}& \red{0.90}& 0.10&0.82&\red{0.80} & 0.20& 0.65& \red{0.90}&0.30&0.70& \red{0.70}&0.05 & 0.30&\red{0.80}&0.05&0.60\\
		FP10 & $\epsilon=2$ &0.80&0.20&\red{0.85}&\red{1.00} & 0.30&0.85& 0.80& 0.10 &\red{1.00}&\red{0.90}&0.40&0.70& \red{0.80}&0.10 & 0.28&0.80&0.14&\red{0.87}\\
		& $\epsilon=5$ &0.80&0.10&\red{0.85}& \red{1.00}& 0.40&0.90& 0.90& 0.10& \red{1.00}&\red{0.90} &0.40&0.70& \red{0.90}& 0.16& 0.30&\red{0.92}&0.20&\red{1.00}\\
		\hline
    	& $\epsilon=0.5$ &0.75&0.21&\red{0.90}&\red{0.90}&0.41&0.88&0.75&0.11&\red{0.90}&0.70&0.30&\red{0.90}&\red{0.77} & 0.10& 0.30&0.55&0.05&\red{0.60}\\
		& $\epsilon=1$  & 0.7&0.15&\red{1.00}& \red{0.90}& 0.40&0.85&0.75 & 0.10& \red{0.80}& 0.80&0.30&\red{0.88}& \red{0.77}& 0.11& 0.30&\red{0.60}&0.05&\red{0.60}\\
		FP20 & $\epsilon=2$  &0.75 &0.15&\red{1.00}& \red{0.95} & 0.55&0.85& 0.85&0.05 &\red{1.00} & \red{0.80}&0.50&0.76& \red{0.70}& 0.20& 0.28&0.80&0.15&\red{0.87}\\
		& $\epsilon=5$  &0.75 &0.35&\red{1.00}&\red{1.00} & 0.50&0.85& 0.90& 0.30& \red{1.00}& \red{0.90}&0.45&0.75& \red{0.80}&0.20 & 0.30&0.90&0.25&\red{1.00}\\
		\hline 
		& $\epsilon=0.5$ &0.95&0.18&\red{1.00}&0.80& 0.60&\red{1.00}&\red{0.90}&0.15&\red{0.90}&0.80&0.45& \red{1.00}& \red{0.80}& 0.10& 0.25&\red{0.58}&0.12&0.53\\
		& $\epsilon=1$  & \red{1.00}&0.12&\red{1.00}&  0.80& 0.60&\red{1.00}&\red{0.95} &0.14& 0.90& 0.80&0.42&\red{0.93}&\red{0.80}&0.10 & 0.28&\red{0.58}&0.12&0.40\\
		FP50 & $\epsilon=2$  &\red{1.00} &0.34&\red{1.00} & 0.84& 0.62&\red{0.95}&0.95 &0.26 &\red{1.00} & \red{0.90}&0.40&0.88&\red{0.85}& 0.16& 0.28&0.80&0.27&\red{0.87}\\
		& $\epsilon=5$ & 0.91& 0.362&\red{1.00}& 0.86& 0.62&\red{0.92}&\red{1.00} & 0.26& \red{1.00}& 0.80&0.42&\red{1.00}&\red{0.80}& 0.29& 0.30&\red{1.00}&0.27&\red{1.00}\\
		\hline 
		& $\epsilon=0.5$ &\red{1.00}&0.4&\red{1.00}&0.97& 0.54&\red{1.00}&\red{0.95}&0.24&0.95&0.80&0.49& \red{1.00}& \red{0.90}&0.12 & 0.25&\red{0.68}&0.17&0.53\\
		& $\epsilon=1$  &\red{1.00} &0.37&\red{1.00}& 0.93 & 0.55&\red{1.00}& \red{1.00}& 0.24& \red{1.00}&0.90 &0.53&\red{0.90}& \red{0.90}& 0.15& 0.28&\red{0.58}&0.17&0.4\\
		FP100 & $\epsilon=2$  &\red{1.00} &0.47&\red{1.00} &0.93 &0.67&\red{0.95}&\red{1.00} & 0.26&\red{1.00}& \red{0.89} &0.55&0.88& \red{1.00}&0.20& 0.28&0.77&0.24&\red{0.87}\\
		& $\epsilon=5$  & 0.97&0.49&\red{1.00} &0.89 & 0.67&\red{0.95}&\red{1.00} & 0.37& \red{1.00}&\red{1.00}& 0.54&\red{1.00}& \red{1.00}& 0.30& 0.30&0.96&0.30&\red{1.00}\\
		\hline
		& $\epsilon=0.5$ &\red{1.00}&0.5&\red{1.00}&\red{1.00}& 0.65&\red{1.00}&\red{1.00}&0.31&\red{1.00}&\red{1.00}&0.70& \red{1.00}& \red{0.90}& 0.23& 0.25&\red{0.78}&0.24&0.53\\
		& $\epsilon=1$  &\red{1.00} &0.57& \red{1.00}& \red{1.00}& 0.69&\red{1.00}&\red{1.00} &0.36& \red{1.00}&\red{1.00}&0.70&0.808& \red{1.00}& 0.23& 0.28&\red{0.71}&0.27&\red{0.71}\\
		FP200 & $\epsilon=2$  &\red{1.00} &0.56&\red{1.00}&\red{1.00} & 0.82&0.95&\red{1.00} & 0.37& \red{1.00}&\red{1.00} &0.68&0.80 & \red{1.00}&0.31& 0.30&0.85&0.30&\red{0.87}\\
		& $\epsilon=5$  & 0.97&0.55&\red{1.00} & \red{1.00}& 0.82&0.95&\red{1.00} & 0.38&\red{1.00} &\red{1.00} &0.68&0.85& \red{1.00}&0.35 & 0.30&\red{1.00}&0.34&\red{1.00}\\
		\hline
		& $\epsilon=0.5$ &0.910&0.750&\red{0.410}&0.908& 0.701&\red{0.300}&0.909&0.819&\red{0.431}&0.899&0.810& \red{0.240}& 0.909& 0.852& \red{0.470}&0.868&0.856&\red{0.540}\\
		& $\epsilon=1$  &0.929 &0.752&\red{0.310} &0.861 & 0.759&\red{0.360} &0.963 & 0.828&\red{0.203}& 0.900&0.806&\red{0.304}& 0.908& 0.850& \red{0.510}&0.849&0.851&\red{0.480}\\
		JSD & $\epsilon=2$  &0.912 &0.754&\red{0.320} &0.824& 0.762&\red{0.399} &0.953 & 0.829& \red{0.202}&0.873 &0.807&\red{0.348}& 0.899& 0.843& \red{0.53}&0.910&0.804&\red{0.500}\\
		length & $\epsilon=5$ & 0.886& 0.753 &\red{0.335} &0.769& 0.760&\red{0.402} &0.936& 0.828& \red{0.234}& 0.828&0.807&\red{0.377}& 0.890&0.804 & \red{0.520}&0.855&0.800&\red{0.530}\\
		\hline 
		& $\epsilon=0.5$ &2311&4002&\red{2299}&3356& 2910&\red{1771}&2299&3002&\red{1813}&2999&2639& \red{1389}& \red{5199}& 6988& 5675&5799&8085&\red{5760}\\
		EMD & $\epsilon=1$  &2219 &4112&\red{1766} & 3438 &2718 &\red{1682}& 2263& 2995& \red{1543}&2923  &2637&\red{1259}& \red{5109}& 6509& 5777&5652&7998&\red{5638}\\
		src-dst & $\epsilon=2$  & 2278&3166&\red{1690}  &3470 &1859 &\red{1673}& 2272& 2874& \red{1356}& 2899 &2545&\red{1284}&4809 & 6444& \red{4697}&5032&7503&\red{4769}\\
		(meters) & $\epsilon=5$  &2354 &3127& \red{1613}&3137& 1835&\red{1499}& 2242& 2885& \red{1295}&2801&2555&\red{1219}&4557 & 5985& \red{4392}&4841&7100&\red{4588}\\
		\hline 
		& $\epsilon=0.5$ &\red{199}&1333&699&\red{40}& 1303&1002&\red{55}&706&621&689&887& \red{670}& \red{680}& 2003& 2082&2174&2002&\red{1990}\\
		EMD& $\epsilon=1$  & \red{154}&1251&709 & \red{425}& 1308&808&\red{46} & 783& 650& 480&769&\red{429}&\red{405} &1850 & 2081&2185&1999&\red{1909}\\
		 density& $\epsilon=2$  & \red{147}&1087&797 &\red{471} & 1260&732& \red{92}& 737& 510& \red{394}&782&409& \red{380}&1300 & 987&1233&1677&\red{1189}\\
		 (meters) & $\epsilon=5$  &\red{179} &1083&741 &\red{451} & 734&788& \red{135}&1300 & 490& \red{357}&723&488&\red{353} &1114 &1174 &1150& 1588& \red{1290}\\
		\hline
    \end{tabular}}
    \end{small}}
    
    \end{table}
\end{landscape}

\pagebreak

\bibliographystyle{ACM-Reference-Format}
\bibliography{bibliography}

\appendix

\section{Additional results}
\label{app:complexity}
\noindent \textbf{Trace Generation Complexity:}
  We compute the number of generation steps of a single synthetic trajectory for each scheme as follows. Let us assume that all generative models are precalculated and saved. For a trace $t$ with length $|t|$, \our takes $Z_{TI}+Z_{TPG}(K)+K^2\log K + 2\cdot|t|$ steps, where $Z_{TI}$ and $Z_{TPG}(K)$ are the respective steps of the TI and TPG models, $K^2\log K$ comes from Dijkstra`s algorithm, $2\cdot|t|$ from the MH and looping parts of the model, and $K$ is the number of top-K locations. As the graphs generated by TPI are precalculated, $Z_{TPG}(K)$ is ignored in he rest of the analysis. Since $|t| = O(K)$,  the number of \our's generative steps is dominated by Dijkstra`s algorithm, thus its complexity is $O(K^2\log K)$.
	
	AdaTrace's generation steps depend on two random walks started from the start and destination points. The total number of generation steps accumulates into $2\cdot m\cdot l$, where $m$ is the size of the grid, and $l$ is the maximal size of the generated trace ($l$ is drawn from a distribution calculated from the original lengths). Since $l\leq m$, AdaTrace's complexity is $O(m^2)$. If the grid is an order of magnitude larger than the number of top-K locations in \our, the empirical number of steps of AdaTrace can be larger than $K^2\log K + Z_{TI} + 2\cdot|t|$, e.g. $m=K^2$. 
	
	The expected number of steps in Ngram depends on the underlying data distribution. Evaluating Ngram on the taxi datasets, the average length of the generated traces are 3 steps only.
	
	Experiments were conducted using Tensorflow 2.0 and Python 3.6.9 on a single Linux server with 98GB RAM and 16 cores. Running time is heavily dependent on the size of the input dataset. For the largest, the SF dataset the generation time for Ngram is on average 1 minute, for AdaTrace is 20 second. In case of \our the trainings of the neural networks take up a lot of time. The overall running time of \our is approximately 3 hours. However, due to the fact that we are considering offline models, it only has to be done once.

	\begin{figure*}[!h]
\begin{center}
        \begin{subfigure}[]{0.60\textwidth}
                \includegraphics[width=\textwidth]{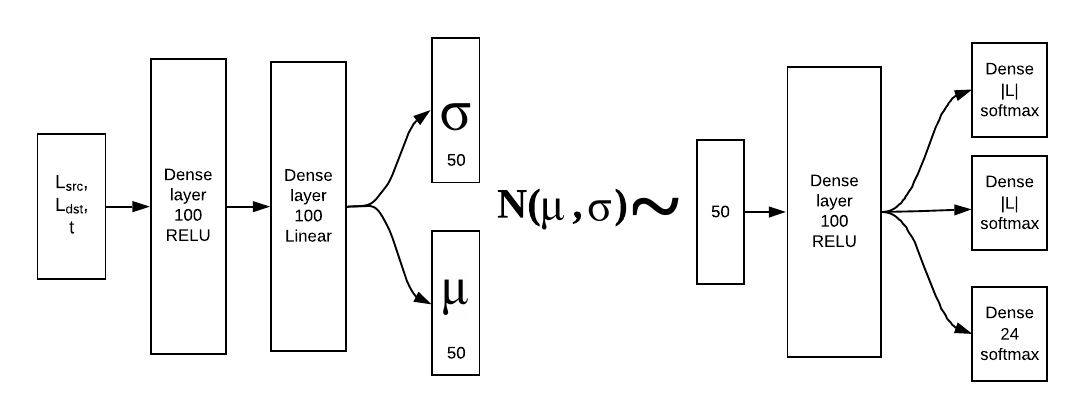}
                \caption{Trajectory Initialization model}
                \label{fig:vae}
        \end{subfigure}
        \begin{subfigure}[]{0.60\textwidth}
                \includegraphics[width=\textwidth]{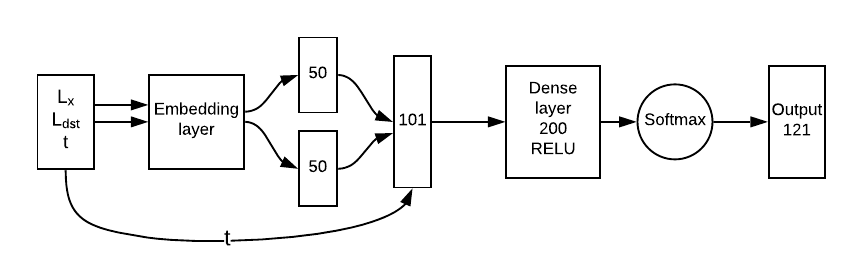}
                \caption{Transition Probability Generator model}
                \label{fig:tpg}
        \end{subfigure}
        \caption{The neural network architectures used in \our}
    \label{fig:models}
    \end{center}
\end{figure*}

	\begin{figure*}[!h]
        \centering
        \begin{subfigure}[]{0.3\textwidth}
                \includegraphics[width=\textwidth]{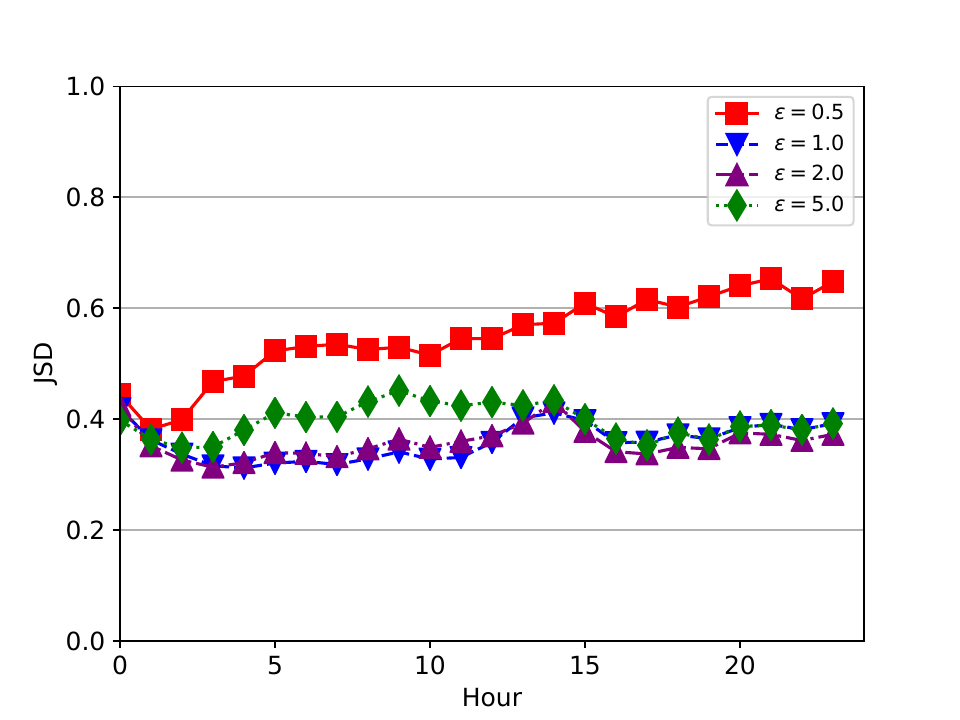}
                \caption{JSD, cell size: 250 m}
                \label{fig:jsd250sf}
        \end{subfigure}
          \vspace{-0.08cm}
        \begin{subfigure}[]{0.3\textwidth}
                \includegraphics[width=\textwidth]{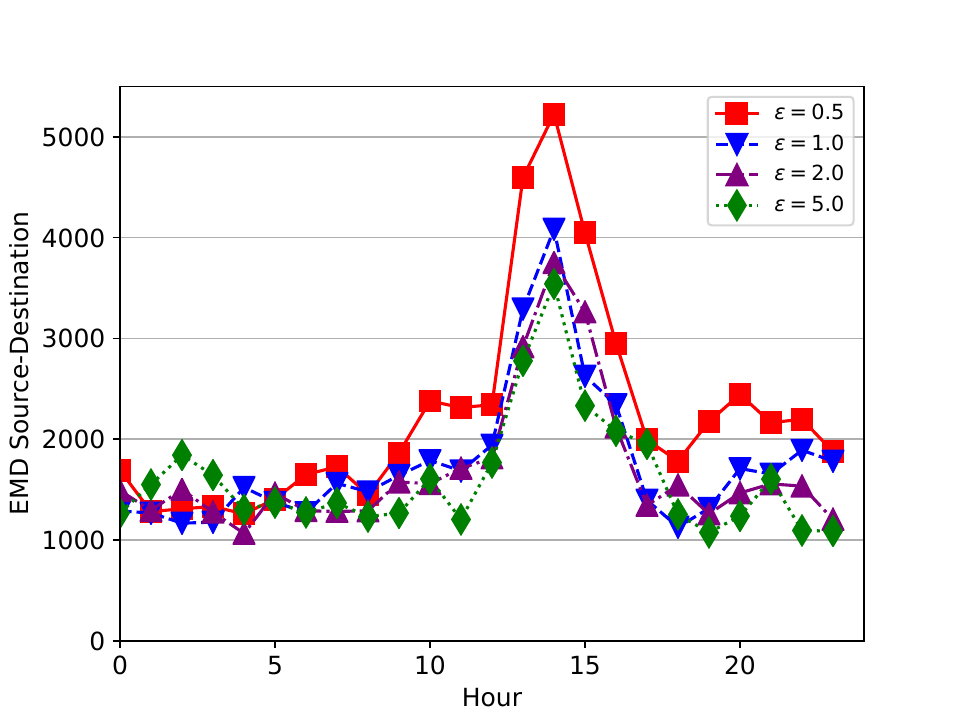}
                \caption{EMD-SD (VAE), cell size: 250 m}
                \label{fig:emdvae250sf}
        \end{subfigure}
          \begin{subfigure}[]{0.3\textwidth}
                \includegraphics[width=\textwidth]{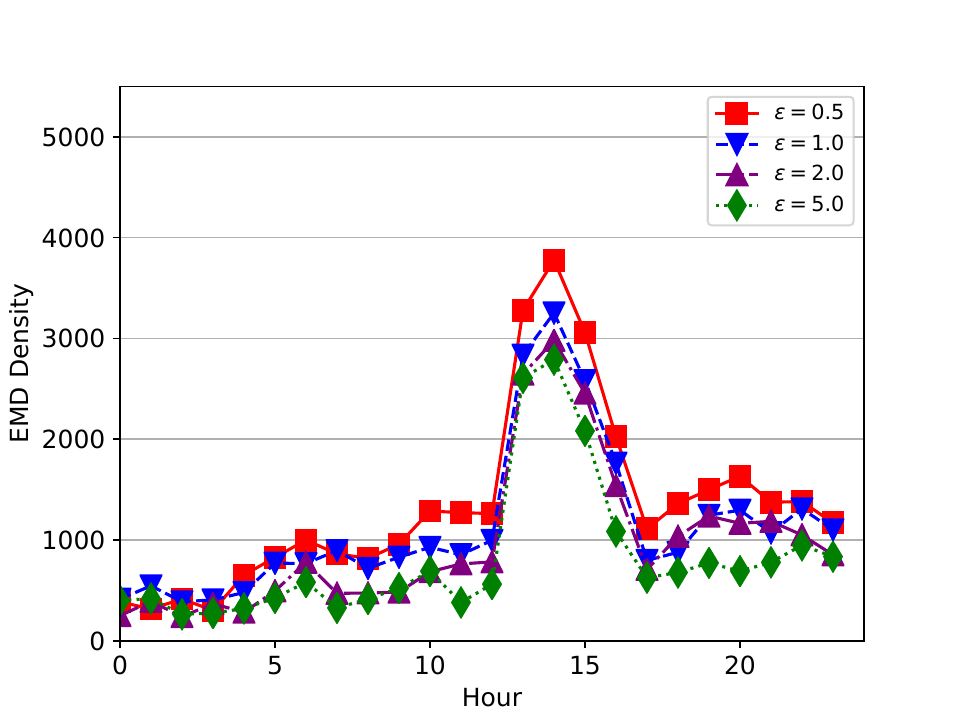}
                \caption{EMD-Density, cell size: 250 m}
                \label{fig:emddens250sf}
        \end{subfigure}
         \begin{subfigure}[]{0.3\textwidth}
                \includegraphics[width=\textwidth]{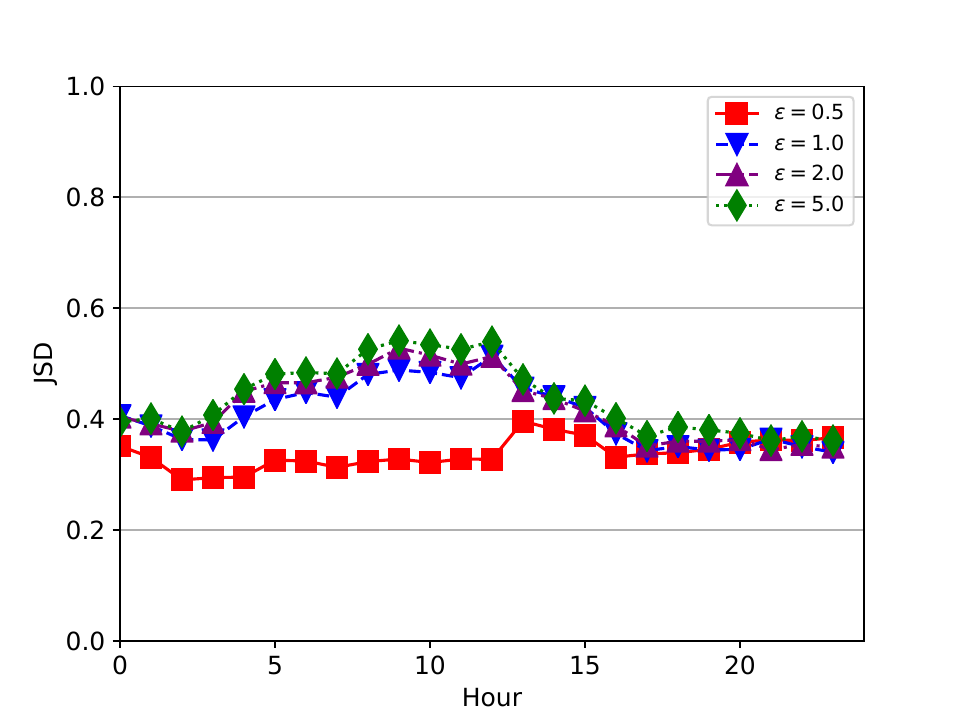}
                \caption{JSD, cell size: 500 $m$}
                \label{fig:jsd500sf}
        \end{subfigure}
        \begin{subfigure}[]{0.3\textwidth}
                \includegraphics[width=\textwidth]{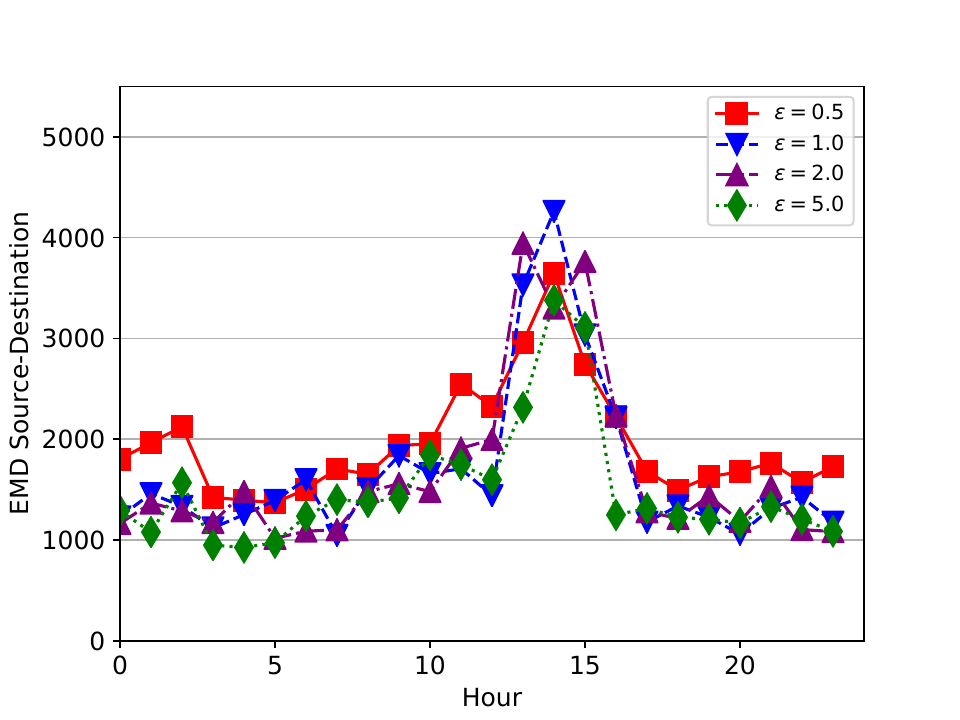}
                \caption{EMD-SD (VAE), cell size: 500 $m$}
                \label{fig:emdvae500sf}
        \end{subfigure}
    \begin{subfigure}[]{0.3\textwidth}
                \includegraphics[width=\textwidth]{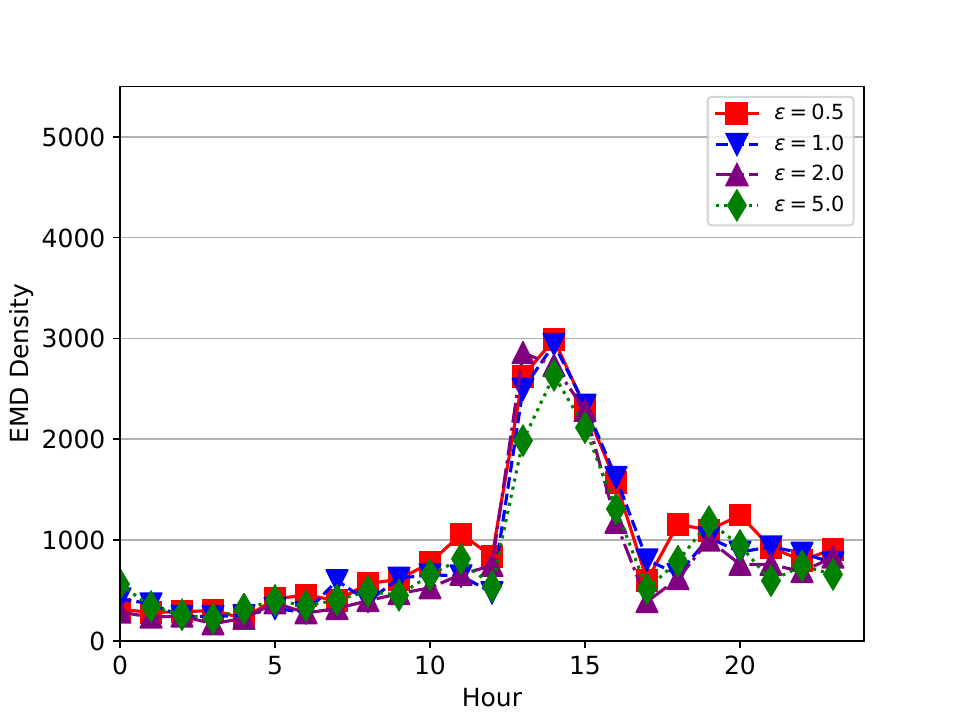}
                \caption{EMD-Density, cell size: 500 $m$}
                \label{fig:emddens500sf}
        \end{subfigure}
        \caption{Performance of our approach on San Francisco dataset depending on the time ($\delta=4\cdot10^{-6}$).}
    \label{fig:perf_on_time_sf}
\end{figure*}

\end{document}